\documentclass[11pt,sort&compress]{elsarticle}

\usepackage[utf8]{inputenc}
\usepackage[margin=1in]{geometry}
\usepackage{graphicx}
\usepackage{multirow}
\usepackage{amssymb}
\usepackage{amsmath}
\usepackage{setspace}
\usepackage{outlines}
\usepackage{enumitem}
\usepackage{xcolor}
\usepackage{upgreek}
\usepackage{mathabx}
\usepackage{algorithm}
\usepackage{algorithmic}
\usepackage{amsthm}
\usepackage[labelfont=bf,font=small]{caption}
\usepackage{epsfig}
\usepackage{geometry}
\usepackage{subfigure}
\usepackage[textsize=tiny]{todonotes}
\usepackage[normalem]{ulem}
\usepackage{lipsum}
\usepackage{array}
\usepackage{booktabs}
\usepackage{lineno}
\usepackage{array}
\usepackage{tikz}
\usepackage[english]{babel}

\usepackage[]{siunitx}
\usepackage[
    protrusion=true,
    activate={true,nocompatibility},
    final,
    tracking=true,
    kerning=true,
    spacing=true,
    factor=1100]{microtype}
    
\SetTracking{encoding={*}, shape=sc}{40}

\usepackage{outlines}
\usepackage{enumitem}

\definecolor{lightblue}{rgb}{0.63, 0.74, 0.78}
\definecolor{seagreen}{rgb}{0.18, 0.42, 0.41}
\definecolor{orange}{rgb}{0.85, 0.55, 0.13}
\definecolor{silver}{rgb}{0.69, 0.67, 0.66}
\definecolor{rust}{rgb}{0.72, 0.26, 0.06}
\definecolor{purp}{RGB}{68, 14, 156}

\definecolor{zblue}{RGB}{8,81,156}
\definecolor{zpurp}{RGB}{84,39,143}
\definecolor{zred}{RGB}{165,15,21}

\colorlet{lightrust}{rust!50!white}
\colorlet{lightorange}{orange!25!white}
\colorlet{lightlightblue}{lightblue}
\colorlet{lightsilver}{silver!30!white}
\colorlet{darkorange}{orange!75!black}
\colorlet{darksilver}{silver!65!black}
\colorlet{darklightblue}{lightblue!65!black}
\colorlet{darkrust}{rust!85!black}
\colorlet{darkseagreen}{seagreen!85!black}

\usepackage{hyperref}
\hypersetup{
  colorlinks=true,
}

\usepackage[nameinlink]{cleveref}
\crefname{equation}{}{}
\def\appendixname{}
\crefname{appendix}{}{}

\usepackage{setspace}
\usepackage{soul}
\sethlcolor{yellow}

\usepackage[parfill]{parskip}

\usepackage{bm}

\newcommand{\Dm}[1]{\ensuremath{\boldsymbol{\varphi} } }

\newcommand{\ignore}[1]{}
\newcommand{\fracp}[2]{\frac{\partial #1}{\partial #2}}

\renewcommand{\[}{\left[}
\renewcommand{\]}{\right]}
\renewcommand{\(}{\left(}
\renewcommand{\)}{\right)}

\newcommand{\cC}{\mathcal{C}}
\newcommand{\Rdot}{\dot{R}}

\newcommand{\bb}{\mathrm{b}}
\newcommand{\dd}{\mathrm{d}}

\newcommand{\mm}{\mathrm{m}}
\newcommand{\vv}{\mathrm{v}}

\newcommand{\Ca}{\mathrm{Ca}}
\newcommand{\We}{\mathrm{We}}
\newcommand{\De}{\mathrm{De}}
\newcommand{\Rey}{\mathrm{Re}}

\newcommand{\testF}{\Delta\psi_0}

\newcommand{\tcollapse}{{t}_{\rm c}} 

\newcommand{\Rmax}{R_{\max}}

\begin{document}

\hypersetup{
  linkcolor=darkrust,
  citecolor=seagreen,
  urlcolor=darkrust,
  pdfauthor=author,
}

\begin{frontmatter}

\title{{\bf\Large Parsimonious inertial cavitation rheometry via bubble collapse time}}

\author[inst1]{Zhiren~Zhu$^*$}
\author[inst2]{Sawyer~Remillard$^*$}
\author[inst1]{Bachir~A.~Abeid}
\author[inst3]{Danila~Frolkin}
\author[inst4,inst5,inst6]{Spencer~H.~Bryngelson}
\author[inst3,inst7]{Jin~Yang}
\author[inst2]{Mauro~Rodriguez~Jr.}
\author[inst1]{Jonathan~B.~Estrada}

\affiliation[inst1]{organization={Department of Mechanical Engineering},
            addressline={University of Michigan},
            city={Ann Arbor},
            postcode={48105},
            state={MI},
            country={USA}}

\affiliation[inst2]{organization={School of Engineering},
            addressline={Brown University},
            city={Providence},
            postcode={02912},
            state={RI},
            country={USA}}

\affiliation[inst3]{organization={Department of Aerospace Engineering and Engineering Mechanics},
            addressline={The University of Texas at Austin},
            city={Austin},
            postcode={78712},
            state={TX},
            country={USA}}

\affiliation[inst7]{organization={Texas Materials Institute},
            addressline={The University of Texas at Austin},
            city={Austin},
            postcode={78712},
            state={TX},
            country={USA}}

\affiliation[inst4]{organization={School of Computational Science and Engineering},
            addressline={Georgia Institute of Technology},
            city={Atlanta},
            postcode={30332},
            state={GA},
            country={USA}}

\affiliation[inst5]{organization={Daniel Guggenheim School of Aerospace Engineering},
            addressline={Georgia Institute of Technology},
            city={Atlanta},
            postcode={30332},
            state={GA},
            country={USA}
            }

\affiliation[inst6]{organization={George W.\ Woodruff School of Mechanical Engineering},
            addressline={Georgia Institute of Technology},
            city={Atlanta},
            postcode={30332},
            state={GA},
            country={USA}}


\begin{abstract}
The rapid and accurate characterization of soft, viscoelastic materials at high strain rates is of interest in biological and engineering applications.
Examples include assessing the extent of tissue ablation during histotripsy procedures and developing injury criteria for the mitigation of blast injuries. 
The inertial microcavitation rheometry technique (IMR, Estrada et al.,~2018) allows for the characterization of local viscoelastic properties at strain rates up to \SI{1E8}s$^{-1}$.
However, IMR now typically relies on bright-field videography of a sufficiently translucent sample at $\geq$1 million frames per second and a simulation-dependent fit optimization process that can require hours of post-processing.
Here, we present an improved IMR-style technique, called parsimonious inertial microcavitation rheometry (pIMR), that parsimoniously characterizes surrounding viscoelastic materials.
The pIMR approach uses experimental advancements to estimate the time to first collapse of the laser-induced cavity within approximately \SI{20}{\nano\second} and a theoretical energy balance analysis that yields an approximate collapse time based on the material viscoelasticity parameters.
The pIMR method closely matches the accuracy of the original IMR procedure while decreasing the computational cost from hours to seconds while potentially reducing reliance on ultra-high-speed videography.
This technique can enable nearly real-time characterization of soft, viscoelastic hydrogels and biological materials with a numerical criterion assessing the correct choice of model. 
We illustrate the efficacy of the technique on batches of tens of experiments for both soft hydrogels and fluids. 

\end{abstract}


\end{frontmatter}

\section{Introduction}\label{sec:intro}

The characterization of viscoelastic soft materials undergoing fast, finite deformations is necessary for a wide range of applications.
These include, but are not limited to, the prediction of biological tissue damage due to blunt impact and blast events~\cite{estrada-etal_2021,wright-etal_2013}, the design of acoustically-responsive scaffolds for drug delivery~\cite{abeid_2024}, and the modeling of non-invasive laser-~\cite{vogel-etal_1996} and ultrasound-based surgical procedures~\cite{bailey-etal_2003,xu-etal_2005,parsons-etal_2006}.
Notably, the United States Food and Drug Administration (FDA) recently approved the clinical treatment of liver cancer with histotripsy, a novel technique that ablates diseased tissues with ultrasound-induced cavitation~\cite{vidal-etal_2022,mendiratta-etal_2024}.
However, soft materials such as hydrogels are challenging to characterize due to their low elastic shear modulus, which ranges from \SI{100}{\pascal} to \SI{1}{\mega\pascal}, and the difficulties of gripping and manipulating the specimens during experiments.
To characterize materials with high compliance and, correspondingly, slow shear wave speed, traditional high-strain-rate experiments, such as the Kolsky bar, must be supplemented with pulse shaping, weak signal sensing, and/or other complicating techniques~\cite{chen_2016}.
Furthermore, soft biological tissues often exhibit spatial heterogeneity, increasing the difficulty of measuring their material property distribution with conventional methods that only provide a macroscale average modulus.
Bio-inspired material systems fabricated to reproduce these functional gradients are similarly difficult to characterize. 

The aforementioned challenges necessitate a technique to locally assess the high strain rate and finite deformation behavior of soft materials. 
Crosby~et~al.\ first developed needle-induced cavitation rheology as an approach to probe the local elastic properties of soft materials~\cite{zimberlin-etal_2007, kundu-crosby_2009}.
A cavity of air or immiscible liquid is injected into the characterized media.
The elastic modulus is determined from the pressure and bubble radius at the onset of mechanical instability.
This quasi-static approach has been extended in recent years to a ballistic strain-rate regime of approximately \qty[mode = text]{1e4}{s^{-1}} by \citet{milner-hutchens_2019,milner-hutchens_2021_multicrack}.
Cohen and co-workers introduced the capability to cyclically expand and relax the needle-induced cavity at controlled stretch rates~\cite{raayai-cohen_2019,chockalingam-etal_2020}, enabling finite deformation characterization of viscoelastic materials.
The inertial microcavitation rheometry (IMR) technique, introduced by \citet{estrada-etal_2018} and improved by others recently~\cite{yang-etal_2020,spratt-etal_2021,mcghee-etal_2023}, accesses a higher range of strain rates by using laser-induced cavitation (LIC) in soft, hydrated materials (i.e., with shear moduli below $\sim$\SI{1}{\mega\pascal}).
An ultra-high-speed camera images the bubble kinematics and the viscoelastic properties of the cavitated media are inversely characterized according to an inertial cavitation bubble model~\cite{keller-miksis_1980, prosperetti-lezzi_1986} with refinements accounting for a two-component mixture of bubble content with heat diffusion and mass transfer~\cite{akhatov-etal_2001,nigmatulin-etal_1981, prosperetti-etal_1988,barajas-johnsen_2017} and stress field in the surrounding media~\cite{yang-church_2005,hua-johnsen_2013,warnez-johnsen_2015,gaudron-etal_2015}.

IMR inversely characterizes viscoelasticity at strain rates reaching \SI{1E3}-\SI{1E8}s$^{-1}$ but has only been successfully applied to characterize nearly transparent materials using ultra-high-speed imaging (rates above 270,000 frames per second). 
The reliance on the full dynamics acquired from high-speed, bright-field videography of the cavity restricts assessment to experimental systems that produce accurate bubble images, which itself is a product of, e.g., camera sensitivity and the material turbidity. 
Increasing the exposure time to combat low light throughput works against the maximum frame rate, thus suggesting a need for us to reduce the reliance on the transient dynamics for the characterization of challenging optical systems. 

Furthermore, the computational cost of the forward simulation, optimization, and best-fit procedure is restrictive, particularly in a potential desired end-case-usage for near on-the-fly characterization. 
Each forward simulation requires about ten seconds.
Batch-fitting multiple experiments simultaneously and increasing the number of model parameters cause an exponential increase in the required forward simulations. 
Hence, we seek to construct an approximate theoretical model that characterizes materials based on just the most essential data drawn from multiple experiments—i.e., maximum radius, quasi-equilibrium radius, and time to first collapse. 
The potential benefit here is twofold: in cases where an approximate (e.g., two-parameter) model is sufficient for predictiveness, the procedure herein represents a rapid characterization method; if accuracy remains critical, this procedure serves to vastly pare down the computational space.
This style of model can also be extended to applications in which the time to collapse is used to quantitatively describe some system behavior or parameters of interest. 
For example, the collapse time measured for LIC in the vicinity of agarose hydrogels was compared against the Rayleigh collapse time (a simplified metric assuming the bubble is just a void) by \citet{sieber-etal_2023} to examine the effect of an elastic boundary. 
\citet{marsh-etal_2024} and \citet{ohl-etal_2006} conducted shock-induced cavitation experiments in water and cervix cell assays, respectively, 
and approximated average velocity and pressure in the resulting jet flow using the Rayleigh collapse time. 
These types of analyses could thus be enhanced by our approximate collapse time model accounting for material behavior and other bubble physics. 

Herein, we use the modified Rayleigh collapse time approach to develop a strategy for the parsimonious characterization of viscoelastic materials that can be described with up to three-parameter models. 
In contrast to prior work~\cite{estrada-etal_2018}, this approach enables the use of data from multiple experiments to arrive at a batch-fit solution. 
The strategy leverages high-fidelity measurements of the maximum bubble radius, the long-term equilibrium bubble radius, and the time from maximum expansion to first bubble collapse.
These quantities of interest are distinctly related through the ultra-high-rate elastic and viscous behaviors of soft materials.
In \Cref{sec:theory+method}, we present an LIC experiment setup capable of quantifying the time of collapse to an accuracy of approximately $\SI{20}{\nano\second}$.
The experiments are complemented with an energy balance analysis that approximately quantifies the effects of material viscoelasticity and secondary factors (viz., surface tension, bubble pressure, and dilatational wave speed) on the time to the first bubble collapse. We then introduce the parsimonious inertial microcavitation rheometry (pIMR) procedure enabled by these experimental and theoretical advancements. 
The consistency of the procedure is verified in \Cref{sec:synthetic_experiment_fitting} with synthetic experiments.
We demonstrate in \Cref{sec:inverse_fit} high-fidelity viscoelastic model parameterization from tens of experiments in viscoelastic liquids and hydrogels, with computational post-processing that takes only seconds. 
In \Cref{sect:discussion}, we discuss the implications of the results obtained and the limitations of the proposed strategy.
We provide concluding remarks in \Cref{sec:conclusion}.

\section{Theory and Methods}
\label{sec:theory+method}
\subsection{Bubble dynamics model}
\label{subsec:bubble_dynamics_model}

We briefly summarize the bubble dynamics model serving as the theoretical basis of the original IMR method. 
A more thorough discussion of the theory, including its underlying assumptions and regimes of applicability, can be found in \citet{estrada-etal_2018}.
The IMR framework has been extensively validated for inertial cavitation in nearly-incompressible, viscoelastic, soft materials ranging from polyacrylamide~\cite{estrada-etal_2018, yang-etal_2020}, agarose~\cite{yang-etal_2022}, and gelatin~\cite{abeid_2024,mcghee-etal_2023} hydrogels to healthy and diseased human liver tissues~\cite{mancia-etal_2019}.

\begin{figure}[tpb]
    \centering
    \includegraphics[scale = 1]{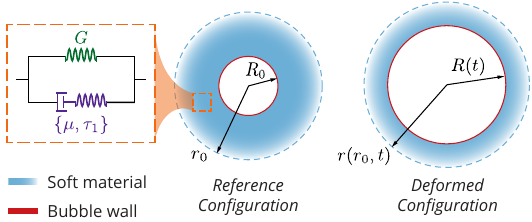}
    \caption{
        Schematic representation of the spherical bubble considered in the bubble dynamics and approximate collapse time models. The nearly incompressible, viscoelastic material surrounding the bubble is modeled as a finite deformation, standard linear solid (SLS)
        described by a ground-state elastic shear modulus $G$, a viscous shear modulus $\mu$, and a relaxation time scale $\tau_1$.
        When $\tau_1\to{0}$, the SLS model becomes a Kelvin--Voigt model.
    }
    \label{fig:bubble-schematics}
\end{figure}

The theoretical model in the IMR framework considers a spherical bubble in an infinite surrounding material environment subjected to a pressure change that causes rapid radial motion, as depicted in \Cref{fig:bubble-schematics}.
The material outside the cavity is viscoelastic and approximated as nearly incompressible.
We denote the equilibrium, stress-free radius of the spherical bubble as $R_0$ and the referential radial coordinate for a material point in the surrounding viscoelastic medium to be $r_0\in[R_0,\infty)$, measured from the center of the bubble to the infinite far field.
Due to the balance of mass, the deformed radial coordinate $r$ and velocity $v$ of a material point $r_0$ at time $t$ in an incompressible medium are
\begin{align}
\begin{aligned}
    r = (r_0^3 + R^3 - R_0^3)^{1/3}, &&
    v = \frac{\dd r}{\dd t}= \dot{R}  \frac{R^2}{r^2},
    \label{eq:incompressible_kinematics}
\end{aligned}
\end{align}
where $R(t)$ is the evolving radius of the bubble.
The balance of linear momentum in the radial direction requires that
\begin{align}
    \rho\(\fracp{v}{t} + v \fracp{v}{r}\) &= - \fracp{p}{r} + \fracp{s_{rr}}{r} + \frac{2}{r}\(s_{rr} - s_{\theta\theta}\) \, ,
    \label{eq:bal-lin-mom}
\end{align}
where $\rho$ is the material density of the surrounding material, $p$ the hydrostatic pressure in the material, $s_{rr}$ and $s_{\theta\theta}$ the normal radial and normal circumferential components of $\bf{s}$, the deviatoric Cauchy stress in the material. 
A perturbation analysis bridging the near- and far-fields of the bubble~\cite{keller-kolodner_1956,keller-miksis_1980,prosperetti-lezzi_1986} leads to a correction of equation \eqref{eq:incompressible_kinematics} accounting for a finite pressure wave speed $c$ in the material and the energy transfer via outward radial acoustic emission. Then, integrating equation \eqref{eq:bal-lin-mom} over $r$ from $r = R$ to $r\to\infty$ results in what is known as the Keller--Miksis equation describing bubble dynamics, 
\begin{align}
   \begin{aligned}
        &
        \(1-\frac{\Rdot}{c}\)R\ddot{R} + \frac{3}{2} \(1-\frac{\Rdot}{3c}\)\Rdot^2 
        &= \frac{1}{\rho}\(1+\frac{\Rdot}{c}\) \(p_\bb - \frac{2\gamma}{R} + S - p_{\infty}\) + \frac{1}{\rho}\frac{R}{c}\dot{\overline{\(p_\bb - \frac{2\gamma}{R} + S \)}},
    \end{aligned}
    \label{eq:keller-miksis}
\end{align}
where overdots denote derivatives with respect to time $t$, 
$p_\bb$ the internal bubble pressure, $p_{\infty}$ the far-field pressure, $\gamma$ the bubble wall surface tension, 
and $S$ the stress integral defined as
\begin{align}
    S &= \int_{R}^{\infty} \frac{2}{r} \(s_{rr} - s_{\theta\theta}\) \dd r.
    \label{eq:stress_intg}
\end{align}

We do not simulate the complex plasma physics contributing to the initial growth of the laser-induced cavity. 
Instead, following a conventional approach for modeling LIC~\cite{nigmatulin-etal_1981,akhatov-etal_2001,barajas-johnsen_2017,estrada-etal_2018,tzoumaka-etal_2023}, we assume that the bubble contents and the surrounding medium reaches thermodynamic equilibrium at maximum bubble expansion and model the bubble dynamics from the instance of maximum bubble expansion.
When considering surrounding material with history-dependent viscoelasticity, we estimate the initial condition of the stress integral $S$ according to a simplified model of the bubble growth phase, as detailed in \ref{sec:synthetic_experiment_fitting} for the Maxwell model.
We assume that the bubble contains a mixture of water vapor and other, non-condensible gas components during the rapid bubble dynamics.
Mass and heat transfer of the two-part bubble contents are assumed to obey Fick's and Fourier's laws, resulting in a set of PDEs~\cite{akhatov-etal_2001, prosperetti-etal_1988, prosperetti_1991}.
Following \citet{estrada-etal_2018}, numerical solutions to the Keller--Miksis equation coupled with the bubble content equations are obtained with the \texttt{ode23tb} function in MATLAB (The MathWorks, Inc., Natick, MA).

\subsubsection{Non-dimensionalization and solution of bubble dynamics model}\label{subsub:nondim}

We follow existing work to non-dimensionalize the governing equations and clarify the interactions between material parameters~\cite{estrada-etal_2018}, and list non-dimensional parameters in \Cref{tab:nondim},
where $\Rmax$ is the maximum radius of the bubble, $G_1$ is the shear modulus associated with the maxwell element, and $v_{\rm c} = \sqrt{p_{\infty}/\rho}$ is the characteristic velocity. 
The non-dimensional Keller--Miksis equation describing the evolution of the non-dimensional bubble radius $R^*$ is
\begin{align}
\begin{aligned}
    \(1-\frac{\Rdot^*}{c^*}\) R^* \ddot{R}^* &+ \frac{3}{2}\(1-\frac{\Rdot^*}{3c^*}\)\Rdot^{*2} = \\
    \(1+\frac{\Rdot^*}{c^*}\) & \(p_\bb^* - \frac{1}{\We \, R^*} + S^* -1\)
    + \frac{R^*}{c^*} \dot{\overline{\(p_\bb^* - \frac{1}{\We \, R^*} + S^*\)}}.
\end{aligned}
\label{eq:nondim_km}
\end{align}
Unless stated otherwise, we assume $\rho = \SI{998.2}{\kilo\gram\per\meter\cubed}$, $p_{\infty} = \SI{101.3}{\kilo\pascal}$, $c = \SI{1484}{\meter\per\second}$, and $\gamma = \SI{0.072}{\newton\per\meter}$.
For the mixture of water and polyethylene glycol (PEG 8000) characterized in \Cref{subsec:PEG_fit}, we assume using the rule of mixtures that $\rho = \SI{1100}{\kilo\gram\per\meter\cubed}$. 
Assuming a constant temperature of \SI{298.15}{\kelvin} in the surrounding material, the material parameters related to the heat and mass transfer of bubble contents are defined according to \citet{estrada-etal_2018}.

\begin{table*}[tb]
    \caption{Dimensionless quantities in the Keller--Miksis Equation.}
    \centering
    \small
    \begin{tabular}{rll} \toprule
        Dimensional quantity             & Dimensionless quantity               & Name \\
        \midrule
        $t$                     &  $t^{*} = t~v_{\rm c}/\Rmax$           &   Time\\
        $R$                     &  $R^{*} = R/\Rmax$                     &   Bubble-wall radius\\
        $R_0$                   &  $R_0^{*} = R_0/\Rmax$                 &   Equilibrium bubble-wall radius\\
        $c$                     &  $c^{*} = c/v_{\rm c}$                    &   Material wave speed\\
        $p_\bb$                 &  $p_\bb^{*} = p_\bb/p_{\infty}$           &   Bubble pressure\\
        $\gamma$                &  $\We = p_{\infty}\Rmax/(2\gamma)$     &   Weber number\\
        $S$                     &  $S^* = S/p_{\infty}$                     &   Stress integral\\
        $G$                     &  $\Ca = p_{\infty}/G$                     &   Cauchy number\\
        $\mu$                   &  $\Rey = \rho v_{\rm c} \Rmax/\mu$     &   Reynolds number\\
        $\tau_1 = \mu/G_1$      &  $\De = \mu v_{\rm c}/(G_1 \Rmax)$     &   Deborah number \\ \bottomrule
    \end{tabular}
    \label{tab:nondim}
\end{table*}

\subsubsection{Stress integral in the surrounding medium}

\begin{table}[tb]
    \caption{Summary of material stress integrals.}
    \renewcommand{\arraystretch}{1.3}
    \centering
    \small
    \begin{tabular}{r l}
    \toprule
        Material model & Stress integral relationship $S^*$ \\ \midrule
        Neo-Hookean
        & $S_{\textrm{NH}}^* = [ 4 ({R_0^*}/{R^*}) + \left({R_0^*}/{R^*}\right)^4 - 5 ] / (2 \, \Ca)$ \\
        Newtonian
        & $S_\vv^* = -(4/\Rey){\Rdot^*}/{R^*}, $ \\
        Kelvin--Voigt
        & $S^*_{\textrm{KV}}=S^*_\vv+S^*_{\rm NH}$\\
        Maxwell
        &  $\De \dot{S}^*_\mm+S^*_\mm = -(4/\Rey){\Rdot^*}/{R^*} $ \\
        Standard linear solid (SLS) &
        $S^*_{\textrm{SLS}} = S^*_\mm + S^*_\mathrm{NH}$ \\ \bottomrule
    \end{tabular}
    \label{table:Stress_int_non_dim}
\end{table}

The stress integral $S^*$ for the viscoelastic constitutive models considered in this work is tabulated in \Cref{table:Stress_int_non_dim}.
We consider finite viscoelasticity constitutive models with stress responses that are additively decomposed into those of three elementary components: a neo-Hookean hyperelastic contribution, a Newtonian viscous contribution, and a Maxwell fading memory viscoelastic contribution.
The stress integral for the Kelvin--Voigt viscoelastic models follow our previous work~\cite{estrada-etal_2018}, in which a neo-Hookean hyperelastic spring is arranged in parallel with a Newtonian viscous dashpot. The standard linear solid (SLS) model (sometimes referred to as the Zener model) consists of a neo-Hookean hyperelastic spring parallel to a Maxwell branch.

Assuming that the characteristic time scale of the bubble oscillation is longer than the time scale of the exponential relaxation of a Maxwell material, its stress integral satisfies 
\begin{align}
    \begin{aligned}
        S^* + \De \dot{S}^* &= -\frac{4}{\Rey} \frac{\Rdot^*}{R^*}.
    \label{eq:linmax}
    \end{aligned}
\end{align}
Due to the fading memory of the Maxwell material, a non-zero stress integral remains at the end of the bubble growth phase, contributing to the ensuing bubble collapse. 
This quantity is numerically evaluated by advancing the ODE \cref{eq:linmax} from the beginning of the growth phase, with initial conditions $R_0$ and $\dot{R} = \dot{R}_i > 0$, with $\dot{R}$ decreasing to $0$ at the end of the growth phase. 
The \texttt{fminsearch} function in MATLAB is used to iteratively solve for $\dot{R}_i$ to minimize $|R - \Rmax|$ at the end of the growth phase. 
The value of $S$ at the end of the growth phase is then determined.
Heat and mass transfer are neglected in these simulations of the bubble growth phase.

These models, or their modified hyperelastic equivalents, have successfully characterized hydrogels with IMR~\cite{estrada-etal_2018, yang-etal_2020, yang-etal_2022, bremer-etal_2024, abeid_2024}. 
In this work, we are primarily interested in the contribution of material viscoelasticity, $S^*$, which in turn influences the collapse time. 
We note that the primary non-dimensional parameters of calibration interest, therefore, are the Cauchy number (ground-state elasticity), $\Ca$, the Reynolds number (ground-state viscosity), $\Rey$, and/or the Deborah number (relaxation time), $\De$, all defined in \Cref{tab:nondim}. 
Thus, in the following sections we distinguish and separately quantify the collapse-time effects from these three material parameters for characterization from those arising from other bubble physics.

\subsection{Energy balance analysis and analytical estimates of collapse time}\label{subsec:energy-budget-framework}

We modify Lord Rayleigh's original analysis to obtain a more accurate prediction of a bubble collapse within hydrogel-like materials. 
Lord Rayleigh utilized an energy balance approach ~\cite{rayleigh_1917} with the following four assumptions: (i) the bubble has no contents, (ii) there is no surface tension between the void and the surrounding material, and the surrounding material is (iii) incompressible and (iv) inviscid.
Thus, the potential and kinetic energy of the surrounding material dictate the evolution of the bubble radius.
Under these conditions, the Keller--Miksis equation \eqref{eq:nondim_km} simplifies to, 
\begin{equation}
    R^*\ddot{R}^* + \frac{3}{2}\Rdot^{*2} = -p_{\infty}^*, 
\end{equation}
where $p_{\infty}^*$ is the non-dimensional liquid pressure ($p_{\infty}^* = 1$).
The potential energy of the inviscid liquid surrounding the bubble is the volume integral of the non-dimensional liquid pressure, 
\begin{gather}
    E^*_\mathrm{LP} = \int_{V^*_b}p_{\infty}^* \dd V^* = p_{\infty}^*V_b^*,
    \label{eq:liq_potential}
\end{gather}
where $V^*_b$ is the volume of the bubble.
The kinetic energy of the liquid is 
\begin{gather}
    E^*_{\textrm{LK}} = \int_{V^*_l}\frac{1}{2} \rho^*u_r^{*2} \dd V^* = \int_{V^*_l}\frac{1}{2}\rho^*\left(\frac{R^{*2}\Rdot^*}{r^{*2}}\right)^2 \dd V^*=2\pi\rho^* R^{*3}\Rdot^{*2}, 
    \label{eq:liq_kinetic}
\end{gather}
where $\rho^*$ is the non-dimensional liquid density ($\rho^* = 1$).
The void is assumed to begin at rest, corresponding to an initial kinetic energy of zero.
The energy balance is then $({4\pi}/{3}) (R^{*3}-1) + 2\pi R^{*3}\Rdot^{*2} = 0$.
Isolating the bubble wall velocity as a function of the radius, $\Rdot^*(R^*(t^*))$, and integrate to the closure of the bubble, we obtain the Rayleigh collapse time
\begin{equation}
    t_{\textrm{RC}}^*= - \int_1^0 \left[-\frac{2}{3}\left(1-\frac{1}{R^{*3}}\right)\right]^{-1/2} \dd R^* = \sqrt{\frac{3 \pi}{2}}\frac{\Gamma [{5}/{6}]}{\Gamma [{1}/{3}]} \approx 0.91468, 
\end{equation}
where $\Gamma[\cdot]$ is the gamma function.
To obtain the dimensional form, we multiply by the characteristic timescale, $\Rmax \sqrt{\rho/p_{\infty}}$.

\subsubsection{General approach for modified Rayleigh collapse time}

We can generalize a Rayleigh-type model as the following equation
\begin{equation}
    R^*\ddot{R}^* + \frac{3}{2}\Rdot^{*2} = -1+ f^*(R^*, \Rdot^*, \ddot{R}^*, S^*, c^*_\mm, R^*_0, p^*_b, \dots),
    \label{eq:genRP}
\end{equation}
where $f^*$ is a sum of different physical phenomena (see \Cref{tab:f_physics}).

\begin{table}[t!]
    \centering
    \small
    \begin{tabular}{r l} \hline
        Phenomenon & Function modifying Rayleigh--Plesset equation $f^*$ \\ \midrule
        Bubble pressure & 
            $f_{bc}^* =p^*_\bb = p^*_\mathrm{go} (R_0/{R^*})^{*~3 \kappa} + p^*_{\vv,\mathrm{sat}}$ \\
        Weak compressibility & 
            $f^*_\mathrm{wc} = ({\Rdot^*}/{c^*_{\mm}}) R^*\ddot{R}^* + ({\Rdot^*}/{2 \, c^*_{\mm}}) \Rdot^{*2} - {\Rdot^*}/{c^*_{\mm}}$ \\ 
        Surface tension & 
            $f^*_{\We} = -1/(\We \, R^*)$ \\ 
        Material response & 
            $f_S^* = S^*$ \\
        Compressibility affecting bubble pressure &   
            $f^*_{cbc} = {\Rdot^*} p_\bb^* / {c_\mm^*} + 
            {R^*}\dot{p}_\bb^*/{c_\mm^*}$ \\
        Compressibility affecting material response & 
            $f^*_{c S} = {\Rdot^*}S^*/{c_\mm^*} +
            {R^*}\dot{S}^* / {c_\mm^*}$ \\ \hline
    \end{tabular}
    \caption{
        Physical phenomena and corresponding functional changes to the Rayleigh--Plesset equation.
        The right-most column sums to the overall function that transforms Eq. \eqref{eq:genRP} into \eqref{eq:nondim_km} (under the polytropic gas assumption). 
    }
    \label{tab:f_physics}
\end{table}%
$f^*$ can also be interpreted as a force or a resistance to force.
Following the work of~\cite{Yang2024}, for a constant $f^*$  an analytical solution for the modified Rayleigh collapse time can be obtained.
Thus, we define a time-averaged $f^*$ acting on the bubble from the surroundings as 
\begin{equation}
    \overline{f}^* = \frac{1}{t_c^*}\int_0^{t_c^*}f^* \, \dd t^* = \frac{1}{t_c^*}\int_1^{0}\frac{f^*}{\Rdot^*} \, \dd R^*.
\end{equation}
Thus, an ansatz for the reduction in the liquid potential energy, and corresponding energy balance are, 
\begin{equation}
    E_f^* = -\frac{4}{3} \pi\overline{f}^* R^{*3},    
\end{equation}
\begin{gather}
    \frac{4}{3}\pi \left(1-\overline{f}^*\right) (R^{*3}-1 ) + 2\pi R^{*3}\Rdot^{*2} = 0,
\end{gather}
respectively. 
Following the procedure of Lord~Rayleigh, the approximate bubble wall velocity is
\begin{equation}
    \Rdot^* \approx -\left[\frac{2}{3}\left(1-\overline{f}^* \right) \left(\frac{1}{R^{*3}}-1\right) \right]^{1/2},
    \label{eq:non-dimensional_approx_BWV}
\end{equation}
with the general approximate modified collapse time
\begin{equation}
    \begin{aligned}
        t_{c}^*& \approx\int_{\Rmax}^0 -\left[-\frac{2}{3} \left(1-\overline{f}^*\right)\left(1-\frac{1}{R^{*3}}\right)\right]^{-1/2} \dd R^* = t_{\textrm{RC}}^*\left(1-\overline{f}^*\right)^{-1/2}.
    \end{aligned}
    \label{eq:general_tc}
\end{equation}
Since time-averaging is a linear operator, we can write the total collapse time modification to be equal to the following, $\overline{f}^* = \sum_{\alpha} \overline{f}^*_{\alpha}$, where $\alpha$ is indexing different physical effects.
We consider the constitutive terms in equation \eqref{eq:nondim_km} individually.
Since inertia dominates the collapse, the interaction of compressibility with other physical phenomena are second-order and are neglected.
Additionally, for simplicity, this analysis will neglect heat and mass transfer in and outside the bubble.
Thus, the vapor pressure in the bubble is constant.

The time averaging of $f^*$ is similar to linearization in traditional perturbation methods. 
That is, if any of the forcing terms approach unity, the approximation will break down and exhibit large errors when compared to the exact solution.
Neo-Hookean elasticity is an exception, as the leading order elastic forcing term is constant.
This exception permits reasonable predictions of the elastic contribution to the collapse time, even for small $\Ca$.

\subsubsection{Bubble pressure effects\label{subsec:bubble_content}}

We assume that there are two primary gases inside the bubble: (1) water vapor and (2) a non-water gas phase.
The latter consists of air and vaporized material that diffuses back into the material over time scales much longer than that of inertial collapse. 
We consider the bubble pressure as the sum of partial pressures of the gases present~\cite{brennen_2014}: $p^*_b = p^*_{\vv}+p^*_{go}( {R^*_0}/{R^*} )^{3\kappa}$, where $\kappa$ the ratio of the heat capacity at constant pressure, $C_P$, to the heat capacity at constant volume, $C_V$.
Additionally, the water vapor pressure is $p^*_{\vv}$, and we assume the non-condensible gas to be polytropic, where $p^*_{go}$ is the equilibrium bubble pressure.

In the limit $R^* \rightarrow 0$ (i.e., infinite bubble pressure), the evaluation of the mean value of the bubble pressure is non-convergent.
Furthermore, there is no expression available for $R_{\min}$ such that we could obtain a finite integrated result.
Since $f^*_{bc} \, \propto \,  p^*_{go}R_0^{*3\kappa}$, then $\overline{f}^*_{bc} \, \propto \,  p^*_{go}R_0^{*3\kappa}$ and a proportionality constant results through integration of $R$ such that
\begin{equation*}
    \overline{f}_{bc}^* = \mathcal{B} p^*_{go} R_0^{*3 \kappa} + p^*_{\rm v},
\end{equation*}
where $\mathcal{B}$ is obtained by numerically solving the exact collapse time integral.
The exact collapse time is found by considering the resistive force of the bubble contents preventing collapse.
The bubble internal energy, or the reduction of liquid potential energy in the presence of bubble contents, is 
\begin{gather}
    E^*_{\mathrm{BIE}} = \int_{V^*_b}-p^*_b dV^* =\int_{V^*_b}-p^*_{go}\left(\frac{V_0^*}{V_b^*}\right)^{\kappa} \dd V^* =  
    \frac{p_b^* \, V_b^*}{\kappa-1}.
\end{gather}
For the special (isothermal) case of $\kappa=1$, the bubble internal energy: 
\begin{gather}
    E_{\mathrm{BIE}}^* = -\frac{4}{3}\pi \, R_0^{*3}\, p_{go}^* \text{log}\left[\frac{4}{3}\pi R^{*3}\right].
\end{gather}
Including the bubble internal energy in the energy balance with the liquid potential  \eqref{eq:liq_potential} and kinetic \eqref{eq:liq_kinetic} energies, the non-dimensional collapse time is
\begin{equation}
    t_c^* = \left\{
    \begin{aligned}
        & \int_1^{R_{\min}^*}-\sqrt{\frac{3}{2}}\left[\frac{1}{R^{*3}}\left(1+\frac{p^*_{\textrm{go}}R_0^{*3 \kappa}}{\kappa-1}-\frac{p^*_{\textrm{go}} R^{*3}}{\kappa -1} \left(\frac{R^*_0}{R^*}\right)^{3\kappa}-R^{*3} \right) \right]^{-1/2} \dd R^*, \quad \text{for} \, \kappa \neq 1, \\
        & \int_1^{R_{\min}^*}-\sqrt{\frac{3}{2}}\left[p_{\rm go}^*R_0^{*3}\frac{\textrm{log}\left(R^{*3} \right)}{R^{*3}}+\frac{1}{R^{*3}}-1\right]^{-1/2} \dd R^*,  \quad \text{for} \, \kappa = 1.
    \end{aligned}
    \right.
    \label{eq:bc_exact_col}
\end{equation}

We evaluate equation \eqref{eq:bc_exact_col} for a small ($0.01$) non-dimensional equilibrium radius.
The integral is evaluated by setting the minimum radius to zero.
Only the real part of the result is considered, which is equivalent to evaluating the integral from $1$ to the minimum radius. 
For the special cases of $\kappa = 1.4$ (isentropic) and $\kappa = 1$, we obtain $\mathcal{B} = 2.1844$ and $1.4942$, respectively.
To be consistent with the initial bubble pressure in IMR~\cite{estrada-etal_2018}, $\kappa = 1$ for the pIMR solver.

\subsubsection{Weak compressibility effects}

In \Cref{tab:f_physics}, the third term of $f^*_\mathrm{wc}$ dominates the weak compressibility effect during the primary bubble collapse.
Thus, $f_\mathrm{wc}^* \approx -M_c \Rdot^*$, where $M_c = 1/c^*_\mm$ is the  characteristic Mach number.
Time averaging $f^*_\mathrm{wc}$ for the duration of the collapse and solving explicitly, 
\begin{gather}
  \overline{f}_\mathrm{wc}^* = \frac{(1-\overline{f}_{c}^*)^{1/2}}{t_{\textrm{RC}}^*} \int_1^{0}-M_c \dd R^* = M_c \frac{(1-\overline{f}_\mathrm{wc}^*)^{1/2}}{ t_{\textrm{RC}}^*} 
  = \frac{2M_c}{M_c+\sqrt{M_c^2+4t_{\textrm{RC}}^{*2}}}.
\end{gather}

\subsubsection{Surface tension effects}

The surface tension of the water-containing material plays a non-negligible role during the collapse.
Time-averaging $f^*_\text{We}$ (see \Cref{tab:f_physics}) we obtain 
\begin{equation}
    \overline{f}_{\We}^* = 
        \frac{(1-\overline{f}_{\We}^*)^{1/2}}{t_{\textrm{RC}}^*} 
        \int_1^{0}\frac{1}{\We \, R^*} 
        \left[\frac{2}{3}\left(1-\overline{f}_{\We^*} \right) \left(\frac{1}{R^{*3}}-1\right) \right]^{-1/2} \dd  R^* = 
        \frac{-\pi}{\sqrt{6} \,  \We \, t_{\textrm{RC}}^*}.
\end{equation}

\subsubsection{Neo-Hookean elasticity}

Approaching bubble closure, $R^*_0 \rightarrow 0$, the neo-Hookean stress integral converges to a constant. 
Therefore, the modification function and collapse time modification are equivalent for a highly inertial collapse, $f^*_{\textrm{NH}} = \overline{f}_{\textrm{NH}}^* = -{5}/{2 \, {\Ca}}$.
Substituting the neo-Hookean expression for $\overline{f}_{\textrm{NH}}^*$ into equation \eqref{eq:general_tc} results in the modified collapse time consistent with the result in \citet{Yang2024}, i.e., 
\begin{equation*}
    t_c^* = t_{\textrm{RC}}^*\left(1+\frac{5}{2 \, \Ca} \right)^{-1/2}.
\end{equation*}
However, for finite $R_0^*$ and $\Ca$ of ${\mathcal{O}}(1)$, the term that is linear in $R_0^*$ can no longer be assumed to be small.
For LIC, $R_0^* \sim 0.1-0.25$, thus we may neglect the quartic term in $R_0^*$ in~\cref{table:Stress_int_non_dim} and $f_{\textrm{NH}}^* \approx (4R_0^*/R^*-5)/2 \, \Ca$.
Accounting for the finite equilibrium bubble radius, the time-averaged $f^*$ is
\begin{equation}
    \overline{f}_{\text{NH}}^* = \frac{1}{t_{\textrm{RC}}^*} 
    \int_1^{0}\frac{1}{2 \Ca}\left[4\frac{R_0^*}{R^*}-5\right]  
    \left[\frac{2}{3} \left(\frac{1}{R^{*3}}-1\right) \right]^{-1/2} \dd R^* = 
    \frac{1}{\Ca}
    \left( \sqrt{\frac{2}{3}} \frac{R_0^* \pi}{t_{\textrm{RC}}^*}-\frac{5}{2}\right).
\end{equation}

\subsubsection{Viscous Newtonian Fluid}
\label{subsubsec:viscous_forcing_function}

Unlike the elastic case, the mean value of the modification function for a viscous Newtonian fluid does not converge. 
For a void to reach closure in a viscous fluid, the strain rate and, therefore, the viscous dissipation becomes infinite.
Similar to the approximation in \cref{subsec:bubble_content}, a proportionality coefficient for $\overline{f}^*_\vv$ is to account for an intractable non-zero minimum radius when evaluating $\overline{f}^*$, i.e., 
\begin{equation}
    \overline{f}^*_{\vv} = \frac{4\left(1-\overline{f}^*_{\vv} \right)^{1/2}}{\Rey \, t^*_{\textrm{RC}}} \int_1^{R^*_{\min}}-\frac{1}{R^*} \dd R^* \approx \frac{-4 \cC \left(1-\overline{f}^*_{\vv} \right)^{1/2}}{t^*_{\textrm{RC}} \, \Rey}.
    \label{eq:viscousfbar}
\end{equation}
Here, $\cC = \cC(\Rey)$ to obtain accurate results at smaller Reynolds numbers which are experimentally relevant.
Solving \cref{eq:viscousfbar} for $\overline{f}_{\vv}^*$ yields
\begin{equation}
    \overline{f}_\vv^* = \frac{4 \cC}{2 \cC+\sqrt{\Rey^2 \, t_{\textrm{RC}}^{*2}+4\cC^2}}.
    \label{eq:viscous_mod_factor}
\end{equation} 
$\cC$ is solved for by balancing the liquid potential \eqref{eq:liq_potential} and kinetic \eqref{eq:liq_kinetic} energies with the viscous dissipation, i.e.,
\begin{gather}
E_{\vv}^* = \frac{16 \pi}{\Rey}\int_0^{t^*} R^*  \Rdot^{*2} \dd t^*.
\end{gather}
The bubble wall velocity and exact collapse time are then
\begin{gather}
    \Rdot^* = -\left[\frac{2}{3}\left(\frac{1}{R^{*3}}-1\right) - \frac{8}{\Rey \, R^{*3}}\int_1^{R^*} R^* \Rdot^* \, \dd R^*  \right]^{1/2},
\end{gather}
\begin{equation}
    t_c^{*}  =
    -\int_1^0\left[\frac{2}{3}
    \left(\frac{1}{R^{*3}}-1\right) - \frac{8}{\Rey \, R^{*3}}\int_1^{R^*} R^* \Rdot^* \, \dd R^* 
    \right]^{-1/2} \dd R^*,
    \label{eq:exact_viscous_collapse}
\end{equation}
respectively.
Substituting Eqs. \eqref{eq:non-dimensional_approx_BWV}, \eqref{eq:general_tc}, and \eqref{eq:viscous_mod_factor} into \cref{eq:exact_viscous_collapse},
the nested integral on the right hand side is evaluated to obtain an implicit relationship for $\cC(\Rey)$,
\begin{equation}
\begin{aligned}
    t_{\textrm{RC}}^*& \left(1-\frac{4 \cC}{2 \cC+\sqrt{\Rey^2t_{\textrm{RC}}^{*2}+4\cC^2}} \right)^{-1/2} = -\sqrt{\frac{3}{2}} \int_1^0\Bigg(-1+\frac{1}{R^{*3}}-
    \frac{4 \sqrt{6}}{\Rey \, \Gamma \left[\frac{5}{3} \right]\, R^{*3}} \Bigg(1\\
    &-\frac{4 \cC}{2 \cC+\sqrt{\Rey^2t_{\textrm{RC}}^{*2}+4\cC^2}}\Bigg)^{1/2}\left[\sqrt{\pi }\, \Gamma\left[\frac{7}{6} \right]\,-2\sqrt{R^*} \, \Gamma\left[\frac{5}{3} \right]\, _2F_1\left( -\frac{1}{2}, \frac{1}{6}; \frac{7}{6}; R^{*3}\right)  \right] \Bigg)^{-1/2} \dd R^*,
    \label{eq:solve_C_visc}
\end{aligned}
\end{equation}
where $_2F_1(\cdot)$ is an ordinary hypergeometric function.
The right-hand side can be numerically integrated to find the value of $\cC$ for a given $\Rey$.
For a fast and simple calculation, we approximate the implicit function with a perturbation series where the small parameter is $1/\Rey$ such that $\cC(\Rey) \approx  \cC_0 + {\cC_1}/{\Rey} + \cC_2 / \Rey^2$.
Constants $\cC_0$, $\cC_1$, and $\cC_2$ approximate the implicit function in equation \eqref{eq:solve_C_visc} and are found by numerically integrating and iterating for three separate Reynolds numbers.
The Reynolds numbers used for this fitting are 18, 25 and 500. 
Below a Reynolds number of 18, the numerical integration produces imaginary solutions.
The following values approximate equation \cref{eq:solve_C_visc} are $\cC_0 = 0.46379$, $\cC_1 = 0.56391$, and $\cC_2 = 5.74916$.

\subsubsection{Kelvin--Voigt viscoelasticity}

The Kelvin--Voigt stress integral average is the sum of the viscous and elastic contributions, i.e.,
\begin{equation}
    \overline{f}_{\text{KV}}^* = \overline{f}_{\vv}^*+\overline{f}_{\textrm{NH}}^*=\frac{4 \cC}{2 \cC+\sqrt{\Rey^2 t_{\textrm{RC}}^{*2}+4\cC^2}} +\frac{1}{\Ca}\left( \sqrt{\frac{2}{3}} \frac{R_0^* \pi}{t_{\textrm{RC}}}-\frac{5}{2}\right).
\end{equation}

\subsubsection{Maxwell viscoelasticity}

The stress integral for the Maxwell model has no closed-form relationship.
We approximate the right-hand side of the stress integral ODE in \Cref{table:Stress_int_non_dim} to be $\overline{f}^*_{\vv}$ and obtain the approximate relationship for the stress integral
\begin{gather}
    S_\mm^*=f_\mm^* \approx\overline{f}_{\vv}^* - \left(\overline{f}_{\vv}^*-f^*_{\rm m,o}\right)\exp\left[ -\frac{t^*}{\De}\right],
    \label{eq:maxwell_stress_evo}
\end{gather}
where $ f^*_{\rm m,o}$ is the unrelaxed stress in the surrounding material at the maximum radius due to the expansion. For $\textrm{De} \ll 1$, $ f^*_{\rm m,o} \approx 0$; otherwise, the initial stress can alter the subsequent bubble dynamics.
The approximation of $ f^*_{\rm m,o}$ is described in section \ref{subsubsec:Maxwell_initial_stress}.
Evaluating the time-average integral yields 
\begin{gather}
    \overline{f}_\mm^* = \overline{f}_{\vv}^*+\frac{\De}{t_c^*}\left(\left(\overline{f}_{\vv}^*-f^*_{\rm m,o}\right)\exp\left[ -\frac{t_c^*}{\De}\right]-\overline{f}_{\vv}^*+f^*_{\rm m,o}\right),
    \label{eq:full_max_fbarstar}
\end{gather}
where $\overline{f}_\mm^*$ is obtained by substituting equation \eqref{eq:general_tc} into the previous expression.
However, the implicit relationship has no analytical solution for $\overline{f}_\mm^*$.
In equation \cref{eq:full_max_fbarstar},
$t_c^*$ is the actual collapse time that depends on $\overline{f}_\mm^*$.
For simplicity, we approximate the remaining collapse time dependence, $t_c^*$, by the Rayleigh collapse time,
\begin{equation}
     \overline{f}_\mm^* = \overline{f}_{\vv}^*+\frac{\De}{t_{\rm RC}^*}\left(\left(\overline{f}_{\vv}^*-f^*_{\rm m,o}\right)\exp\left[ -\frac{t_{\rm RC}^*}{\De}\right]-\overline{f}_{\vv}^*+f^*_{\rm m,o}\right).
    \label{eq:ebf_Maxwell}
\end{equation}

\subsubsection{Standard linear solid using neo-Hookean elasticity}

We consider a material model, the modified standard linear solid or the Zener model \cite{zener1965elasticity,hua-johnsen_2013}, comprising a Maxwell element parallel to a neo-Hookean elastic element to be able to describe more complicated finite-deformation viscoelastic material behavior.
Since the deviatoric Cauchy stress tensor is a sum of contributions, the stress integral and its time derivative are $S_{\textrm{SLS}}^* = S_\mm^*+S_{\textrm{NH}}^*$ and $\dot{S}^*_{\textrm{SLS}} = \dot{S}_\mm^*+\dot{S}_{\textrm{NH}}^*$, respectively.
Similarly, the collapse time modification function is the sum $\overline{f}^*_{\textrm{SLS}} = \overline{f}_{\mm}^*+\overline{f}_{\textrm{NH}}^*$.

\subsubsection{Initial stress due to unrelaxed Maxwell element}
\label{subsubsec:Maxwell_initial_stress}
The energy balance approach is used to approximate the initial, unrelaxed Maxwell stress at the maximum bubble radius.
We approximate the growth process as the inverse of the collapse; therefore, the modification functions $\overline{f}^*$ switch sign.
Additionally, for bubble growth, it is assumed that the fluid is initially stress-free, $f_{\text{m,o}}^*=0$.
Therefore, by using the negative value of $\overline{f}^*_{\vv}$, the growth time, and $f_{\text{m,o}}^*=0$, we approximate the initial Maxwell stress using equation~\eqref{eq:maxwell_stress_evo}. 

To find the growth time, we consider the bubble to be nucleated in a stress-free material at the equilibrium radius with a positive bubble wall velocity such that the correct maximum stretch ratio $R_{\text{max}}/R_0$ is reached.
The non-dimensional energy balance is 
\begin{equation}
     \frac{4}{3}\pi \left(1-\overline{f}_g^*\right) (R_0^{*3}-R^{*3} ) + 2\pi\left(R_0^{*3} \dot{R}_i^{*2}- R^{*3} \dot{R}^{*2}\right) = 0,
     \label{eq:growth_energy_balance}
\end{equation}
where $\dot{R}_i^*$ is the unknown initial bubble wall velocity and $\overline{f}^*_g$ the average of the Rayleigh--Plesset modification function during the growth phase. 
The initial bubble wall velocity is obtained by setting the non-dimensional current bubble radius and bubble wall velocity to 1 and 0, respectively.
Solving for the initial bubble wall velocity yields:
\begin{equation}
    \dot{R}_i^* = \sqrt{\frac{2}{3}\left( 1-\overline{f}^*_g\right) \left(\frac{1}{R^{*3}_0}-1 \right)}.
    \label{eq:growth_radial}
\end{equation}
The current bubble wall velocity is obtained by substituting equation~\eqref{eq:growth_radial} into equation~\eqref{eq:growth_energy_balance} and taking the positive root for bubble growth:
\begin{equation}
    \dot{R}^* = \sqrt{\frac{2}{3}\left( 1-\overline{f}^*_g\right) \left(\frac{1}{R^{*3}}-1 \right)}.
\end{equation}
The growth time is then
\begin{equation}
\begin{aligned}
    t^*_g &=  \int_{R^*_0}^1\left[\frac{2}{3}\left( 1-\overline{f}^*_g\right) \left(\frac{1}{R^{*3}}-1 \right) \right]^{-1/2} dR^*\\
    & = \frac{1}{5\, \Gamma \left[\frac{4}{3} \right]}\sqrt{\frac{6}{1-\overline{f}_g^*}}\left(\sqrt{\pi} \, \Gamma\left[\frac{11}{6}\right]-R^{*5/2}_0 \, \Gamma \left[\frac{4}{3}\right]  \, _2F_1\left( \frac{1}{2}, \frac{5}{6}; \frac{11}{6}; R_0^{*3}\right)  \right).
\end{aligned}
\label{eq:growth_time}
\end{equation}
If $\overline{f}^*_g>1$, then the growth time approximation has an imaginary, unphysical contribution due to averaging the forcing during the growth phase.
The two physical effects that can produce these imaginary solutions are elasticity and surface tension.
Neo-Hookean elasticity and surface tension produce unphysical results for $\text{Ca} < 5/2$ ($G > \SI{50}{\kilo\pascal}$) and We of the same order as $\text{Ca}$, respectively.
The shear moduli observed in this work were all below \SI{25}{\kilo\pascal}. 
For We of $\mathcal{O}(1)$, the maximum radius would be much smaller than is experimentally relevant in this work.
Small Deborah numbers ($<$ 0.1) result in the initial stress prediction deviating from the iterative method result described in \Cref{sec:synthetic_experiment_fitting} with relative errors above 50$\%$ (data not shown).
However, the accuracy of the initial stress in this regime is inconsequential, as the corresponding collapse time modification factor is very small.
For the collapse time prediction, parameter values yielding a $\overline{f}^*>1$ would lead to weak oscillations not inertial bubble collapse.

\subsection{Experimental methods}\label{subsec:experiment_maintext}

\begin{figure}[tpb]
    \centering
    \makebox[\textwidth][c]{
        \includegraphics[scale = 0.75]{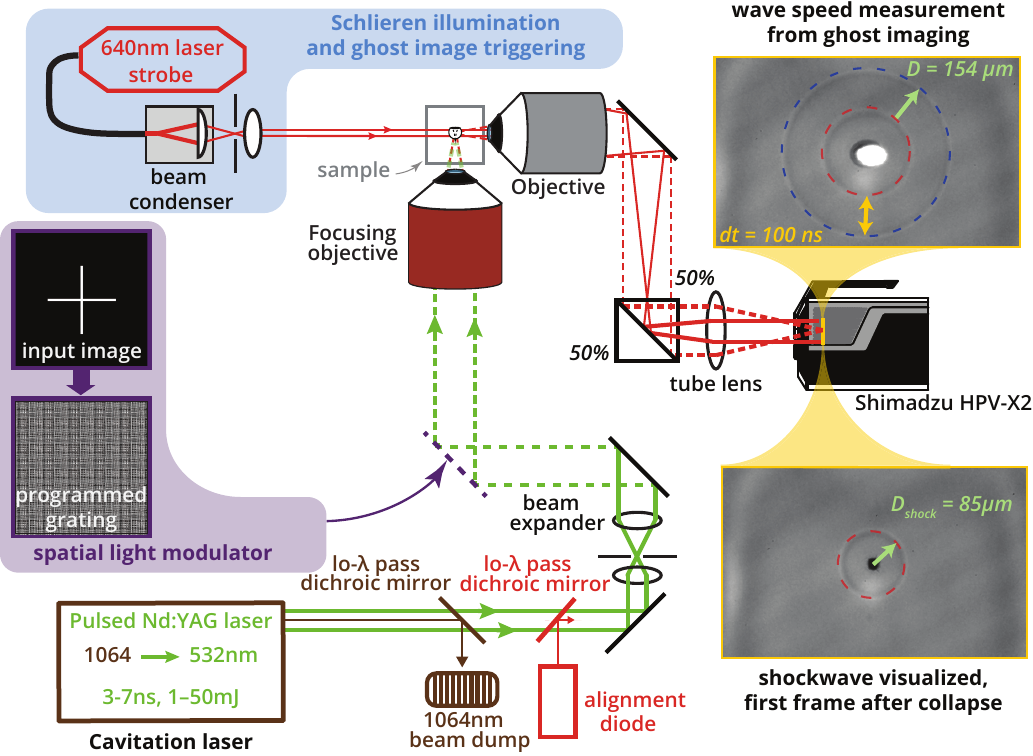}
    }
    \caption{
        The experimental setup to generate, record, and profile single laser-induced microcavitation (LIC) bubble events in soft materials. The setup uses a combination of a class-4, frequency-doubled Q-switched \SI{532}{\nano\meter} Nd:YAG pulsed laser, a high-speed imaging camera, and a spatial light modulator. The time of the first bubble collapse is estimated according to the shock wave, which was visualized by shadowgraph and ghost imaging techniques. 
    }
    \label{fig:Setup}
\end{figure}

The laser microcavitation experiments follow the general LIC procedure of \citet{estrada-etal_2018} with two main advancements of (i) shadowgraph and ghost imaging and (ii) incident beam shaping~\cite{sukovich2020cost} (see \Cref{fig:Setup}).
These are described in detail in \Cref{sec:experiment_detail} and summarized here.

Single LIC bubble events are generated in soft hydrogels using a pulsed, frequency-doubled (\SI{532}{\nano\meter}), Q-switched Nd:YAG laser.
The pulse energy is user-defined and was on the order of $1-$\SI{10}{\milli\joule} for the experiments.
A diffraction-limited focusing objective condenses the laser pulse into a beam waist to approximately \SI{4}{\micro\meter} in diameter.
A second objective is oriented orthogonal to the imaging plane for the purpose of verifying bubble sphericity.
A spatial light modulator is used to tune higher-order beam asymmetry to create spherical bubble events.
We record the microcavitation event at 1 million frames per second (Mfps) using a Shimadzu~HPV-X2 (Tokyo, Japan) ultra-high-speed imaging camera.
A backlight laser strobe fires synchronously with the camera and is sent to the bubble event as parallel light.
This light enables shadowgraph imaging, a mode related to Schlieren imaging that permits the visualization and measurement of emitted shockwaves.
We strobe the shadowgraph backlight twice per frame, improving our estimate of the collapse time using the shock speed (found by locating two shocks on one frame) and the minimum radius estimate.

Polyacrylamide (PAAm) gels for characterization purposes were prepared at concentrations of 5\%/0.03\% and 10\%/0.06\% acrylamide/bisacrylamide (v/v) according to previously developed protocols~\cite{tse2010preparation,estrada-etal_2018}.
The PAAm gels were cast in square \SI{5}{\milli\liter} polystyrene spectrophotometer cuvettes and cured for \SI{45}{\minute} prior to characterization.

Viscous liquid samples were produced by mixing Polyethylene glycol 50\% (w/v) of molecular weight 8,000 (PEG 8000; Avantor 101443-878) with DI water in a proportion of 80\% PEG by volume.
The blended mixtures were poured into glass-bottomed, \SI{35}{\milli\meter} diameter Petri dishes up to roughly \SI{2}{\milli\meter} of depth.
These prepared samples retained a liquid state with no signs of heterogeneity.
The low-frequency shear moduli of the PEG 8000 samples were measured using a TA Instrument ARES-G2 rotational rheometer (New Castle, DE) equipped with a \SI{40}{\milli\meter} diameter stainless steel 2$^{\circ}$-angle cone plate fixture and a flat base. Dynamic loading was applied at a frequency of \SI{1}{\radian\per\second} with the maximum strain amplitude of \SI{0.04}{\radian}.

\subsection{Parsimonious inverse characterization based on bubble collapse time}

\begin{figure}[tpb]
    \centering
    \includegraphics[scale = 1.5]{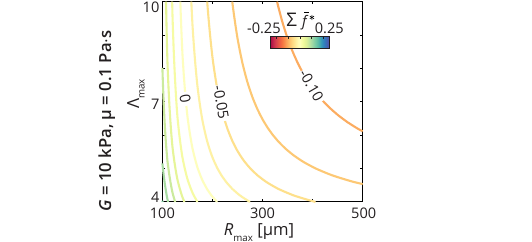}
    \caption{
        Combined effect of viscoelasticity, bubble content pressure, weak compressibility, and surface tension on the bubble collapse time in a Kelvin--Voigt material with $\{ G = \SI{10}{\kilo\pascal}, \mu = \SI{0.1}{\pascal\second} \}$ across typical range of $\Rmax$ and $\Lambda_{\max}$ in LIC experiments.
    }
    \label{fig:Inv-Fit_Example}
\end{figure}

Past studies using IMR have found that adjusting the laser energy can modulate the maximum radius of the bubble,
while the amplification factor of the initial bubble expansion, $\Lambda_{\max} = \Rmax/R_0$, is weakly sensitive to laser energy~\cite{buyukozturk-etal_2022}.
Thus, we perform LIC experiments at various laser energy levels on a material and tune the parameters appearing in equation~\cref{eq:nondim_km}.
Our experiments traverse the $\{\Rmax,\Lambda_{\max}\}$ space for a constant set of dimensional viscoelastic parameters and collect $\tcollapse^*\( \Rmax,\Lambda_{\max} \) $. 
As illustrated in \Cref{fig:Inv-Fit_Example} for a Kelvin--Voigt model example, $\tcollapse^*$ reflects the combined effect of the forcing functions reviewed in $\ref{subsec:energy-budget-framework}$, which varies with $\Rmax$ and $\Lambda_{\max}$.  

We solve for the viscoelastic model parameters that minimize the difference between the collapse times approximated by the energy balance analysis, $\tcollapse^\mathrm{Approx}$, and those that were experimentally measured, $\tcollapse^\mathrm{Expt}$.
We refer to this inverse characterization method as the parsimonious Inertial Microcavitation Rheometry technique (pIMR). 

Specifically, we use a cost function,
\begin{align}
    \psi\[ G, \mu, \tau_1 \] &= \log_{10} \[ \frac{1}{n} \sum_{k=1}^{n} \( \(\frac{\tcollapse^{\rm Expt}{}\[R_{\max{,k}},\Lambda_{\max{,k}}\]}{\tcollapse^{{\rm Approx}}\[ G, \mu, \tau_1, R_{\max,k}, {\Lambda}_{\max,k} \]}\)^2  - 1 \)^2 \],
\end{align}
that quantifies the agreement between the collapse time measured experimentally 
and the approximated value 
for a set of trial parameters $\{ G, \mu, \tau_1 \}$ according to the energy balance analysis using $n$ number of experiments.
The cost function can be interpreted as the order of mean square relative error between the measured and predicted collapse time. 
Using the \texttt{fminsearch} function in MATLAB, which implements a Nelder--Mead direct search process~\cite{nelder-mead_1965,lagarias-etal_1998}, an optimal set of viscoelastic parameters is then determined to minimize $\psi$.

To analyze the precision of the inverse characterization solution, we also introduce the normalized cost function for experimental data,
$
    \hat{\psi} = \psi - \psi_{0}
$,
where $\psi_{0}$ is the minimized cost function for a given type of viscoelastic model, corresponding to the optimal solution found by the direct search algorithm. The normalized cost $\hat{\psi}$ is equal to zero at the optimal solution, whereas the positive-valued $\hat{\psi}$ elsewhere reflects how far the solution is from being optimal.

Furthermore, we seek to encourage the parsimony of the inversely-calibrated viscoelastic model type and minimize the number of parameters used. 
This could be achieved, for example, through a $F$-test-based criterion that discourages the addition of a model parameter that does not lead to a large enough decrement of $\psi_0$. 
In practice, a user could decide to penalize a model multiplicatively based on the added number of terms~\cite{wang-etal_2018,wang-etal_2021,nikolov-etal_2022} or use a least absolute shrinkage and selection operator (LASSO) regression~\cite{tibshirani_1996}. 
In the present work, we simply report cost function decrement $\Delta \psi_0$ to reflect the parsimony of the constitutive model. 
For the specific models we consider, we report the cost function decrements
\begin{align}
    \Delta \psi_0 &=
    \begin{cases}
    \min\[\psi_{0,{\rm NH}},\psi_{0,{\rm Nt}}\] - \psi_{0, {\rm KV}} \, , & \text{Kelvin--Voigt} \\
    \psi_{0, {\rm KV}} - \psi_{0, {\rm SLS}}  \, , & \text{SLS}
    \end{cases}
\end{align}
where $\psi_{0,{\rm NH}}$,~$\psi_{0,{\rm Nt}}$,
~$\psi_{0,{\rm KV}}$,~$\psi_{0,{\rm SLS}}$ are the minimized $\psi$ corresponding to the neo-Hookean, Newtonian, 
Kelvin--Voigt, and SLS models, respectively. 
If a model type results in a decrement below a threshold value of $\Delta\psi_0$, we can consider it to be over-fitting.
In our analysis presented below, a threshold value of $0.5$ is considered for an illustrative purpose.
A user may modify the choice of threshold depending on the type of material characterized and the relative amount of noise in the experiment data. 

\section{Consistency check of pIMR}
\label{sec:synthetic_experiment_fitting}

To check the consistency of the pIMR procedure, we use it to recover input viscoelastic models from synthetic experiments.
Using the bubble dynamics model and stress integral evaluation procedure described in \ref{subsec:bubble_dynamics_model}, we simulate $R(t)$ corresponding to $n = 36$ pairs of $\Rmax \in [100, 400]~\unit{\micro\meter}$ and $\Lambda_{\max} \in [5,9]$. 
The $\Rmax$ and $\Lambda_{\max}$ values are generated with the Latin hypercube sampling method assuming a uniform distribution~\cite{mckay-etal_2000}. 
The simulated collapse time is then used to inversely calibrate viscoelastic models with pIMR. The results are presented in \Cref{tab:synthetic-fit}.

\begin{table*}[tb]
    \caption{Calibrated viscoelastic parameters, minimized cost function, and cost function decrement from synthetic experiments. (Numerical values below \qty{1E-5}{} are reported as $\sim{0}$.)
    }
    \centering
    \small
    \begin{tabular}{c c l l l l l}
        \hline 
        Material & Model & $G$ (\SI{}{\kilo\pascal}) & $\mu$ (\SI{}{\pascal\second}) & $\tau_1$ (\SI{}{\micro\second}) & $\psi_0$ & $\testF$ \\
        \hline 
        Synthetic NH,  & NH  & 10.44 & -- & -- & -6.09 & --  \\
        $G = \qty{10}{\kilo\pascal}$ & Newtonian  & -- & $\sim{0}$ & -- & -1.47 & --  \\
         & KV  & 10.44 & $\sim{0}$ & -- & -6.09 & 0.00 \\
        \cmidrule(lr){2-7}
        Synthetic Newtonian,  & NH  & $\sim{0}$ & -- & -- & -1.83 & --  \\
        $\mu = \qty{0.1}{\pascal\second}$ & Newtonian  & -- & 0.096 & -- & -4.85 & --  \\
         & KV  & $\sim{0}$ & 0.096 & -- & -4.85 & 0.00 \\
        \cmidrule(lr){2-7}
        Synthetic KV,  & NH  & 5.07 & -- & -- & -2.79 & --  \\
        $G = \qty{10}{\kilo\pascal}$,~$\mu = \qty{0.1}{\pascal\second}$ & Newtonian  & -- & $\sim{0}$ & -- & -1.94 & --  \\
         & KV  & 10.11 & 0.095 & -- & -5.50 & 2.71 \\
         & SLS & 10.38 & 0.116 & 0.506 & -5.78 & 0.28 \\
        \cmidrule(lr){2-7}
        Synthetic SLS,  & NH  & 6.56 & -- & -- & -3.41 & --  \\
        $G = \qty{10}{\kilo\pascal}$,~$\mu = \qty{0.1}{\pascal\second}$, & Newtonian  & -- & $\sim{0}$ & -- & -1.80 & --  \\
        $\tau = \qty{1}{\micro\second}$  & KV  & 9.17 & 0.052 & -- & -5.16 & 1.75 \\
         & SLS & 10.08 & 0.096 & 1.90 & -5.97 & 0.81 \\
        \hline 
   \end{tabular}
   \label{tab:synthetic-fit}
\end{table*}

If additional constitutive model parameters are rejected when $\testF < 0.5$, for example, pIMR can correctly identify the type of constitutive model used in the synthetic experiments. 
Compared to the input values, the elastic and viscous shear moduli, $G$ and $\mu$, are recovered to within an accuracy of 5\%. This is well within the confidence interval commonly reported for hydrogels characterized by IMR~\cite{estrada-etal_2018,yang-etal_2020,yang-etal_2022,bremer-etal_2024}, confined and unconfined compression~\cite{strange-oyen_2012,gu-etal_2003,normand-etal_2000}, and indentation~\cite{strange-oyen_2012}.
For the SLS model, the relaxation time scale $\tau_1$ is recovered to within a factor of two of the input value. This accuracy is acceptable since $\tau_1$ contributes to the viscoelastic stress through an exponential relaxation function.

In \ref{sec:more-synthetic-results}, additional synthetic experiments with $n < 36$ and artificially perturbed collapse time data are considered. We find that the accuracy of collapse time measurement in our LIC experiment setup is sufficient to ensure the stable performance of pIMR. Although it is theoretically possible to calibrate a constitutive model with $m$ parameters using data from $n = m$ LIC experiments, this makes the calibration results more susceptible to the inherent discrepancy between the bubble dynamics model and the approximate collapse time model presented in \Cref{subsec:energy-budget-framework}. A larger sample size $n$ is encouraged for the accurate calibration of viscoelastic model type and parameters.

We also verified that the assumption of isothermal bubble content made in \ref{subsec:bubble_content} has a negligible effect on the inverse calibration results. Using collapse time data from synthetic experiments with isothermal bubble content, without heat and mass transfer, pIMR recovered viscoelastic models matching those shown in \Cref{tab:synthetic-fit}. This confirms that the material viscoelasticity is the dominant factor modifying the bubble collapse time in typical LIC experiments.

\section{Inverse characterization of viscoelastic materials}\label{sec:inverse_fit}

The proposed pIMR procedure is applied to the inverse characterization of viscoelastic materials from LIC experiment results. 
The calibrated viscoelastic model parameters are summarized in \Cref{tab:experiment-fit}.

\begin{table*}[tb]
    \caption{Inversely characterized viscoelastic parameters from LIC experiments.}
    \centering
    \small
    \begin{tabular}{l l l l l l l l}
        \hline 
        Material & Technique & Model & $G$ (\SI{}{\kilo\pascal}) & $\mu$ (\SI{}{\pascal\second}) & $\tau_1$ (\SI{}{\micro\second}) & $\psi_0$ & $\testF$ \\
        \hline 
        Water-PEG    & pIMR ($n = 20$) & Newtonian  & -- & 0.223 & -- & -1.72 & -- \\
                     & pIMR ($n = 20$) & KV  & $\sim{0}$ & 0.223 & -- & -1.72 & 0.00 \\
                       & IMR  & Newtonian  & -- & 0.151 & -- &  -- &  -- \\
                       & Shear-plate rheometry  & Newtonian  & -- & 0.122 $\pm$0.005 & -- &  --  & -- \\
                    \cmidrule(lr){2-8}
        PAAm,          & pIMR ($n = 52$) & NH  & 3.11 & -- &  -- & -3.19 & -- \\ %
        5/0.03\%       & pIMR ($n = 52$) & KV  & 6.52 & 0.109 &  --  & -3.28 & 0.09 \\ 
                       & pIMR ($n = 52$) & SLS    & 18.42 & 0.731 & 6.24 & -3.34 & 0.06 \\ 
                       & IMR  & KV  & 5.01 & 0.145 & -- &  -- &  -- \\    
                       & Static compression~\cite{estrada-etal_2018} & NH & 0.461$\pm$0.004 & -- & -- &  -- &  -- \\
                   \cmidrule(lr){2-8}
        PAAm,           & pIMR ($n = 39$) & NH  & 10.11 & -- & -- & -3.18 & -- \\ 
        10/0.06\%       & pIMR ($n = 39$) & KV  & 14.49 & 0.130 & -- & -3.29 & 0.11   \\ 
                       & pIMR ($n = 39$) & SLS    & 21.31 & 0.538 & 6.03 & -3.30 & 0.01  \\ 
                       & IMR  & KV  & 12.02 & 0.115 & --  &  -- &  --  \\    
                       & Static compression~\cite{estrada-etal_2018} & NH & 2.97$\pm$0.06 & -- & -- &  -- &  --   \\
        \hline 
   \end{tabular}
   \label{tab:experiment-fit}
\end{table*}

\subsection{Characterization of water-PEG mixture (viscous fluid)}
\label{subsec:PEG_fit}

\begin{figure}[tpb]
    \centering
    \includegraphics[scale = 1.5]{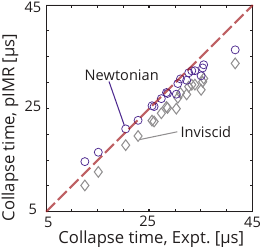}
    \caption{
    Comparison of measured vs. predicted collapse time for the 80\% (v/v) concentration water-PEG mixture characterized, with $n = 20$ samples. The collapse time during LIC is larger than the predicted value for an inviscid material, suggesting the dominance of viscosity over elasticity during the collapse process.
    }
    \label{fig:EBF-vs-Expt_Peg}
\end{figure}

We characterized uncured mixtures of water and PEG 8000, with an 80\% v/v PEG concentration. 
The mixtures are expected to exhibit viscous fluid behavior with negligible elasticity.

A total of $20$ LIC experiments were performed, with $\Rmax$ ranging from $\qty{103.7}{\micro\meter}$ to $\qty{342.9}{\micro\meter}$ and $\Lambda_{\max}$ ranging from $4.73$ to $7.10$.
The inverse fitting of the Kelvin--Voigt model 
converged to a Newtonian model with minimal elasticity.
\Cref{fig:EBF-vs-Expt_Peg} shows the approximated collapse time $\tcollapse^{\rm Approx}$ for the calibrated model versus the measured collapse time $\tcollapse^{\rm Expt}$ of each experiment.
We observe that $\tcollapse^{\rm Expt}$ is larger than the predicted value for an inviscid material, confirming the dominance of material viscosity over elasticity during the bubble collapse.
If the water-PEG material were Newtonian, the pIMR and experimentally observed collapse times would fall along the $y=x$ line.
However, the Newtonian pIMR model overpredicts the collapse time for experimental collapse times less than $\SI{25}{\micro\second}$, and underpredicts for longer times. 
Therefore, we hypothesize that the fluid is exhibiting non-Newtonian behavior in the high-strain rate regime.

\begin{figure}[tpb]
    \centering
    \includegraphics[scale = 1.5]{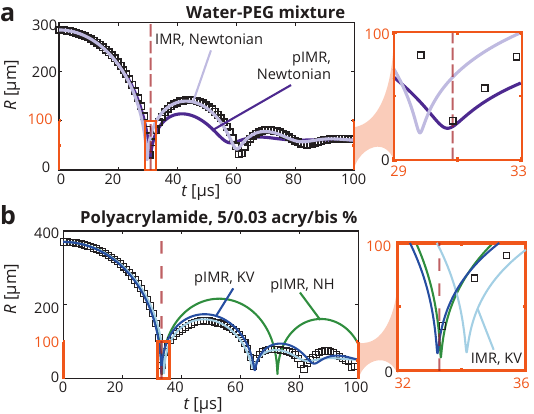}
    \caption{
        Bubble dynamics corresponding to representative experiment data (hollow squares) and the inverse characterization solutions: (a) water-PEG 8000 mixture, (b) PAAm gel with 5/0.03\% (v/v) acrylamide/bisacrylamide concentration. 
    }
    \label{fig:Experiment-Bubble-Dynamics}
\end{figure}

Using the IMR technique, a Kelvin--Voigt model with $\mu = \SI{0.151}{\pascal\second}$ and $G = \SI{0}{\pascal}$ (i.e., a Newtonian model) was found to minimize the offset between the normalized bubble history $\{t^*,R^*(t^*)\}$ recorded experimentally and simulated by the bubble dynamics model, up to the third oscillation peak. 
\Cref{fig:Experiment-Bubble-Dynamics}~(a) shows $R(t)$ for a typical experiment with the simulated bubble dynamics and the inversely characterized constitutive model parameters.
As we expect, the optimal parameters of a Newtonian model calculated via pIMR produce dynamics that closely match the bubble collapse time, with an error of \SI{0.29}{\micro\second} (relative error: $0.94\%$).
The IMR-calibrated model reproduced the post-collapse bubble dynamics more accurately than pIMR but failed to capture the correct collapse time (see~\cref{fig:EBF-vs-Expt_Peg} frame a).

The least-squares fitting method employed by IMR 
obtains agreement between simulation and experimental data of the entire transient bubble dynamics. 
As a result, individual time events, such as the primary bubble collapse time, can be inaccurate.
Thus, to accurately reproduce the post-collapse dynamics of non-Newtonian fluids (see \Cref{fig:EBF-vs-Expt_Peg}), shear-dependent viscosity models, e.g., Carreau model, are  needed for pIMR.
While this behavior is not the focus of this work, it does  warrant further investigation.

\subsection{Characterization of polyacrylamide gels}

We characterize PAAm gels with two different concentrations of acrylamide/bisacrylamide.
This class of material has been characterized with IMR in past studies~\cite{estrada-etal_2018,tzoumaka-etal_2023} and exhibited viscoelastic behaviors that were captured well by the Kelvin--Voigt model.

A total of $52$ LIC experiments were performed on specimens with an acrylamide/bisacrylamide concentration of 5/0.03\% (v/v), with $\Rmax = \SIrange{218.0}{401.3}{\micro\meter}$ and $\Lambda_{\max}=\SIrange{6.49}{8.46}{}$.
The history of $R(t)$ for a representative experiment is shown in \Cref{fig:Experiment-Bubble-Dynamics}~(b) with the simulated bubble dynamics of the calibrated models.
The bubble dynamics of the pIMR neo-Hookean and Kelvin--Voigt solutions matched the collapse time within \SI{0.087}{\micro\second} (relative error: $0.26\%$)
and \SI{0.063}{\micro\second} (relative error: $0.19\%$), respectively. The IMR-calibrated Kelvin--Voigt model overestimates the collapse time by \SI{0.84}{\micro\second} (relative error: $2.5\%$). 
If we were to reject a constitutive model when $\testF < 0.5$, for example, the calculated $\testF$ suggests that the neo-Hookean model suffices to describe the scaling of bubble collapse time. However, the Kelvin--Voigt model clearly reproduces the post-collapse bubble dynamics more accurately in \Cref{fig:Experiment-Bubble-Dynamics}~(b). 

A total of $39$ LIC experiments were performed on gels with an acrylamide/bisacrylamide concentration of $10/0.06\%$ (v/v),
with $\Rmax = \SIrange{215.2}{416.3}{\micro\meter}$ and $\Lambda_{\max}=\SIrange{5.67}{6.76}{}$. Again, cost function decrement $\testF$ suggests that the collapse time scaling is described well by the neo-Hookean model. Consistent with the IMR calibration results, pIMR suggests that the elastic modulus increases with the concentration of acrylamide/bisacrylamide, while the viscosity changes minimally between the two types of specimens.

\section{Discussion}\label{sect:discussion}

The inverse characterization results in \Cref{sec:inverse_fit} show pIMR estimating finite deformation viscoelastic model parameters across a batch of experiments with different $\Rmax$ and $\Lambda_{\max}$.
For the $52$-sample batch of PAAm gel (5/0.03\% (v/v) acryl/bis) experiments, the optimal model type and parameters for all samples were determined within $1$ second of computation on a workstation (Intel~Core~i7~14700K). 
Using the IMR bubble dynamics model, approximately $10$ seconds of computational time are required to simulate the bubble dynamics up to the fourth peak of oscillation for each set of input parameters describing the material viscoelasticity and the bubble's initial and equilibrium conditions in each experiment.
The computational cost is amplified as the simulation is repeated for combinations of input parameters. 

The estimation of viscoelastic properties from collapse time also reduces the requirements that IMR previously placed on the optical turbidity of the characterized material.
With a decreased frame rate and an increased exposure time per frame, bright-field videography can be used to measure the maximum and equilibrium radii of the bubble in an optically turbid material.
Since the bubble collapse coincides with the emission of shock waves, its occurrence can be captured with methods other than the optical strategy introduced in \Cref{subsec:experiment_maintext}. 
Integrated circuit piezoelectric transducers are commonly used in shock tube~\cite{bentil-etal_2016,salzar-etal_2017,marsh-etal_2024} and Kolsky bar~\cite{chen_2016,saraf-etal_2007} experiments to detect pressure spikes during high strain rate deformation of materials.
Past studies of laser- and ultrasound-induced cavitation have used hydrophones to acquire acoustic signals and identify the occurrences of shockwave-emitting collapse events~\cite{akhatov-etal_2001,sukovich-etal_2016}.
Using custom-built histotripsy arrays with receive capable elements, Sukovich et al. have demonstrated the experimental quantification of the time lapse between the nucleation and the first collapse of ultrasound-induced cavitation in ex-vivo porcine and bovine tissues~\cite{macoskey-etal_2018,sukovich-etal_2020,haskell-etal_2025}. These acoustic techniques can be feasibly integrated into a LIC experiment setup for the inverse characterization of optically turbid materials.

\begin{figure}[tpb]
    \centering
    \includegraphics[width=\textwidth]{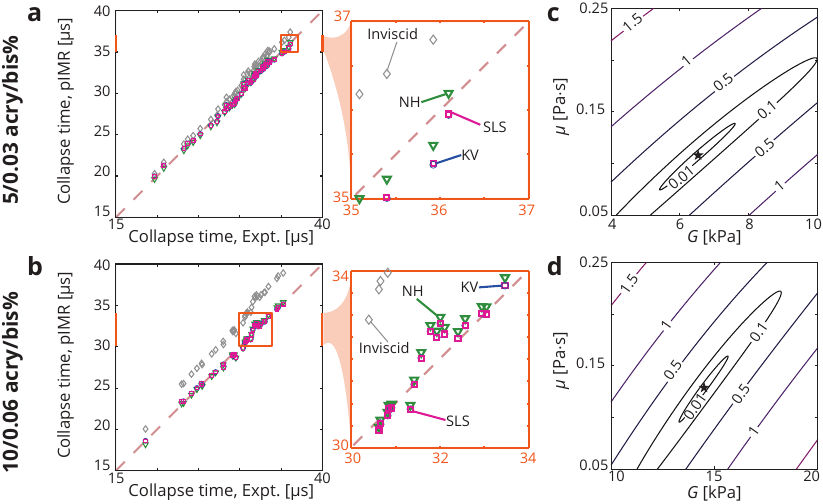}
    \caption{
        Characterization of PAAm gels with pIMR.
        Comparison of measured vs. predicted collapse time for (a) 5/0.03\% (v/v) acrylamide/bisacrylamide and (b) 10/0.06\% (v/v) acrylamide/bisacrylamide for inviscid, neo-Hookean, Kelvin--Voigt, and Standard Linear Solid models. 
        Contours of normalized cost function $\hat{\psi}$ corresponding to Kelvin--Voigt parameters $\{G,\mu\}$ for (c) 5/0.03\% (v/v) acrylamide/bisacrylamide and (b) 10/0.06\% (v/v) acrylamide/bisacrylamide experiments. The opposing effects of elastic modulus $G$ and viscous modulus $\mu$ on the bubble collapse time are reflected in the slope of the $\hat{\psi}$ space.
    }
    \label{fig:PA-Fits}
\end{figure}

The inverse characterization of PAAm gels suggest that, at the length scale of LIC experiments, the material elasticity has a stronger contribution to the bubble collapse time than the material viscosity. This agrees with past studies concluding that the first collapse of LIC in hydrogels are dominated by inertial and elastic effects~\cite{yang-etal_2020,Yang2024}.
In \cref{fig:PA-Fits}~(a) and (b), the approximated collapse times $\tcollapse^{\rm Approx}$ for calibrated models are plotted against the measured collapse time $\tcollapse^{\rm Expt}$ of each experiments in the two types of PAAm gels.
For each LIC experiment, $\tcollapse^{\rm Expt}$ is shorter than what is predicted for an inviscid fluid, indicating that the ground-state elasticity is more dominant than the material viscosity during the bubble collapse.
As shown in \Cref{tab:experiment-fit}, the neo-Hookean model sufficiently decreases $\psi_0$ and seems to be the optimal choice of constitutive model for the PAAm gel.
However, a close inspection of \Cref{fig:PA-Fits}~(a) and (b) reveals that the calibrated neo-Hookean model underestimated the collapse time in experiments with small $\Rmax$ and overestimated the collapse time in experiments with large $\Rmax$.
The addition of material viscosity improved the agreement between the measured and predicted collapse time values since $\overline{f}^*_\vv$ increases in magnitude with smaller $\Rey$, which is inversely proportional to $\Rmax$ when $\mu$ is constant.

The LIC experiments of PAAm gels surveyed ranges of $\Rmax$ with a ratio of $\sim2$ between the upper and lower bounds. The Kelvin--Voigt parameters, $G$ and $\mu$, calibrated with pIMR have relative errors within 30\% of the IMR result.
However, the normalized cost function spaces shown in \Cref{fig:PA-Fits}~(c) and (d) reflect a lower precision in the calibration of $\mu$ compared to $G$. 
For example, in the case of the 10/0.06\% (v/v) acrylamide/bisacrylamide experiments, the region of $\hat{\psi} < 0.1$ spans $\qty{11.1}{\kilo\pascal} \leq G \leq \qty{18.2}{\kilo\pascal}$ and $\qty{0.03}{\pascal\second} \leq \mu \leq \qty{0.22}{\pascal\second}$, corresponding to upper-to-lower bound ratios of $1.6$ and $7.3$, respectively.
The precision of the calibrated viscosity can be improved by surveying as broad of a range of bubble sizes as is experimentally feasible. 
As discussed in \ref{subsubsec:viscous_forcing_function}, the viscous forcing function $\overline{f}^*_{\vv}$ depends on $\Rey$, which is linearly proportional to $\Rmax$. 
Maximizing the experimental range of $\Rmax$ would distinguish this effect of material viscosity on the bubble collapse time.
In our current setup, practical experimental considerations bound the maximum (in this case, due to the finite cuvette size) and minimum values (e.g., due to the camera frame rate) of $\Rmax$. 
The range of $\Rmax$ for LIC experiments could, in general, be broadened with longer-focal-length objectives permitting larger cuvettes/samples and maximum bubble sizes, or sub-nanosecond laser pulses to produce more reliable small bubble events. 
In the case of more complicated design spaces than $\Rmax$ alone (such as a variety of external pressures or initial stretch ratios), the pIMR method could be performed in tandem with a recently developed Bayesian optimized experimental design procedure by \citet{chu-etal_2025} currently only employing bubble dynamics forward simulations. 
In this scenario, the pIMR approach further stands to speed up characterization by using the analytical model for informing the next best experiment to run for maximum information gain. 

In addition to the initial collapse time considered in the present work, other measurable parameters or constitutive models with additional effects (e.g., non-Newtonian behavior) may be harnessed for a more effective parsimonious characterization of viscoelastic models.
For example, \Cref{fig:Experiment-Bubble-Dynamics}~(b) shows that the neo-Hookean and Kelvin--Voigt solutions from pIMR lead to bubble dynamics that diverge more discernibly from each other after the initial collapse.
Similarly, for the case of a water-PEG mixture, shown in \Cref{fig:Experiment-Bubble-Dynamics}~(a), the difference between the Newtonian fluid models calibrated via IMR and pIMR produced becomes clearer in the post-collapse bubble dynamics.
Comparing the pIMR and IMR diverging solutions past the primary bubble collapse, a hybridized approach containing the speed of pIMR and accuracy of IMR is warranted. 
That is, pIMR is used to identify the constitutive model and estimates of the associated material parameters for an accurate and fast IMR inverse characterization.

\section{Conclusion}
\label{sec:conclusion}

We present the pIMR technique, a parsimonious enhancement of the IMR technique that rapidly characterizes the local viscoelastic properties of soft materials from laser-induced cavitation experiments.
This new procedure is possible due to experimental advancements in estimating the collapse time of a laser-induced cavity, coupled with a theoretical energy balance analysis. 
We make an ansatz to a modified potential energy through averaging effects within the Keller--Miksis equation.
This ansatz allows the collapse time approximation to include viscoelastic parameters, surface tension, bubble pressure, and finite wave speed.
In our approach, we do not introduce empirical fitting parameters in the energy balance analysis to improve its agreement with the bubble dynamics model. 
These approximate models for the collapse time were shown to perform well in predicting the collapse time from simulations of the Keller--Miksis equation over a parameter space that is experimentally relevant to inertial microcavitation within soft materials.

The proposed procedure successfully pares down the space upon which we seek the global optima of viscoelastic model parameters.
Using a cost function $\psi$ that quantifies the agreement between the measured and predicted collapse time, our procedure identifies the simplest type of constitutive model and the optimal values of model parameters.
Experimental characterization of viscous fluid and hydrogel specimens resulted in optimized Newtonian and Kelvin--Voigt parameters, respectively, that closely matched the results of the IMR procedure while reducing the computational cost of post-processing 
from more than an hour to a few seconds. 

Our LIC experiments in viscous fluids and soft hydrogels revealed that the dominating mechanisms during the first collapse of the bubble do not necessarily dominate during the ensuing bubble dynamics. 
For the case of PAAm gels, a neo-Hookean hyperelastic model suffices to reproduce the bubble collapse, while the post-collapse kinematics is strongly influenced by the material viscosity and described better by a Kelvin--Voigt model.
We envision that this issue can be addressed via an inverse characterization procedure considering additional observable parameters in the post-collapse bubble dynamics, physics in the models (e.g., non-Newtonian behavior), or through coupling pIMR with IMR.

While the present work only considers viscoelastic models with three or fewer parameters in part due to potential non-uniqueness of solutions involving collapse time alone, the procedure of finding the corresponding modification functions can be straightforwardly extended to viscoelastic models with other non-linear elastic and non-Newtonian fluid behaviors. 
For example, by recording data over a range of stretch ratios as in \cite{buyukozturk-etal_2022} and incorporating higher-order elastic spring elements \cite{yang-etal_2020}, the non-linear elastic response can be decoupled from the shear modulus. 
Quantification of higher-order non-linear behavior is a potentially impactful future use case of this method, as an apparent assessment need in histotripsy is the quantification of the degree of therapy completion via acoustic emissions~\cite{sukovich-etal_2020,haskell-etal_2025}. 
\citet{haskell-etal_2025} found an increase in the time from bubble initiation to collapse during the course of therapy, which is qualitatively due to conversion of elastic biomaterial to an ablated viscous liquid. 
The method presented herein could quantitatively describe the material mechanics during the course of the therapy, and hence, the time to therapy completion. 
We thus anticipate pIMR to be a useful tool in establishing mechanics-based therapy guidelines for different prospective tissue applications. 

\section*{Conflict of Interest}

The authors have no known conflicts of interest associated with this publication, and there has been no significant financial support for this work that could have influenced its outcome.

\section*{Data availability}

{
The code for pIMR is available at: \url{https://github.com/InertialMicrocavitationRheometry/parsimonious_IMR}.
}

\section*{Acknowledgments}

SHB and JBE acknowledge support from the U.S. Department of Defense, the Army Research Office under Grant No. W911NF-23-10324 (PMs Drs.\ Denise Ford and Robert Martin). 
MRJ acknowledges support from the U.S. Department of Defense under the DEPSCoR program Award No. FA9550-23-1-0485 (PM Dr. Timothy Bentley).
JBE, MRJ, and JY gratefully acknowledge support from the U.S. National Science Foundation (NSF) under Grant Nos. 2232426, 2232427, and 2232428, respectively.

\bibliographystyle{bibsty.bst}
\bibliography{mybibfile}

\appendix

\section{Experimental methods}\label{sec:experiment_detail}

Our setup generates, records, and profiles pulses of single LIC bubble events in soft materials using a combination of a pulsed, Q-switched, user-adjustable \SIrange{1}{25}{\milli\joule}, frequency-doubled \SI{532}{\nano\meter} Nd:YAG laser (Continuum Minilite II, San Jose, CA) and a high-speed imaging camera (HPV-X2; Shimadzu, Kyoto, Japan).
The setup is triggered using an 8-channel pulse/delay generator (Model 577; Berkeley Nucleonics, San Rafael, CA) according to a customized pre-programmed pulse sequence.
The pulse sequence was validated using an oscilloscope (P2025; Berkely Nucleonics).
Sequential triggering signals fire two single pulses: the first triggers the laser’s flash lamp, and the second fires the Q-switch.
The last two triggering signals are sent to a beam profiler (BC106N-VIS; Thorlabs) and the high-speed camera.
The backside of the sample is illuminated with the aid of a \SI{640}{\nano\meter} monochromatic ultra-high-speed strobed diode laser (Cavilux Smart UHS; Cavitar, Tampere, Finland).
The high-speed camera sync-out signal triggers the illumination laser.
The laser beam/pulse was aligned to the back apparatus of a 10X/0.25 High-Power MicroSpot Focusing Objective (LMH-10X-532; Thorlabs, Newton, NJ) using three reflective broadband dielectric mirrors (BB1-E02; Thorlabs, Newton, NJ), three short-pass dichroic mirrors, a beam-sampler, and a spatial light modulator (SLM) (Holoeye, Berlin, Germany).
The first dichroic mirror (DMSP605; Thorlabs, Newton, NJ) is used for the beam alignment in conjunction with a continuous exposure Collimated Laser Diode Module (CPS635R; Thorlabs, Newton, NJ).
A 2X fixed magnification beam-expander (GBE02-A; Thorlabs, Newton, NJ) helps distribute the collimated beam on a larger area and minimizes any potential damage to the SLM and focusing objective lens at the back aperture.
The second high-pass dichroic mirror (DMSP550; Thorlabs, Newton, NJ), which has a cutoff wavelength of 550 nm, was used to filter infrared wavelengths and discard them into a beam-block (LB2; Thorlabs, Newton, NJ).
The visible beam is then reflected onto a spatial light modulator, allowing for higher control over the last pulse shape and energy.
Last, the beam is split before it reaches the focusing objective using a beam sampler lens (BSF10-A; Thorlabs, Newton, NJ).
Approximately $0.5\%$ of the split beam is reflected towards a beam profiler (BC106N-VIS; Thorlabs, Newton, NJ) to assess the pulse quality and measure its energy.
The remaining $99.5\%$ of the beam continues to the focusing objective through the third dichroic mirror (DMSP550; Thorlabs, Newton, NJ), which also has a cutoff wavelength of \SI{550}{\nano\meter}, allowing the cavitation laser (\SI{532}{\nano\meter}) to pass while reflecting the illumination laser light (\SI{640}{\nano\meter}).
The focusing objective focuses the beam at the microcavitation imaging plane.

The microcavitation event is performed at 1 million frames per second (Mfps) using a Shimadzu HPV-X2 (Tokyo, Japan) high-speed imaging camera, illuminated by CAVILUX Smart UHS (Tampere, Finland) laser, and through both, the cavitation objective and an Olympus Plan 10X-0.25 Achromat imaging objective (RMS10X; Thorlabs, Newton, NJ). The data is analyzed using our in-house Matlab image processing code.
To measure the wave speed in the medium, we deployed two imaging techniques simultaneously: laser shadowgraph~\cite{traldi2018schlieren} and ghost imaging~\cite{agrevz2020high}.
Shadowgraph imaging is performed by manipulating the backlighting path to capture density variation due to the compressive shockwave.
The physical location of the pressure wave is then estimated during the bubble's cavitation and collapse.
Ghost imaging is achieved by triggering the strobed backlight a user-defined number of times per camera exposure, usually $2$ or $3$ per frame.

\section{Additional consistency checks of pIMR}
\label{sec:more-synthetic-results}

In this section, we present additional synthetic experiments to check the consistency of pIMR. We consider a Kelvin--Voigt material with $G = \qty{10}{\kilo\pascal}$ and $\mu = \qty{0.1}{\pascal\second}$.

\begin{figure}[tpb]
    \centering
    \includegraphics[scale = 1]{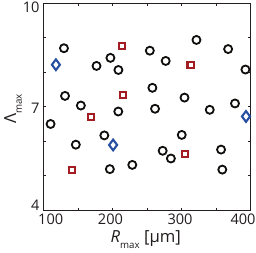}
    \caption{
    Distribution of $\{\Rmax,\Lambda_{\max}\}$ in synthetic experiments. 
    All data points are considered in the $n = 36$ case. 
    The red square and bubble diamond points are considered in the $n = 9$ case. 
    Only the blue diamond points are considered in the $n = 3$ case.
    }
    \label{fig:synthetic-pool}
\end{figure}

As discussed in \Cref{sec:synthetic_experiment_fitting}, when the synthetically generated collapse time for $n = 36$ combinations of $\Rmax \in [100, 400]~\unit{\micro\meter}$ and $\Lambda_{\max} \in [5,9]$ are considered, pIMR identifies the Kelvin--Voigt model to be the optimal choice and recovers $G$ and $\mu$ to within an accuracy of $5\%$. 
To examine the effect of the sample size $n$, we alternatively consider subsets with $n = 9$ and $n = 3$, as illustrated in \Cref{fig:synthetic-pool}. 
The corresponding results from pIMR are shown in \Cref{tab:extra-synthetic-fit}. 
The $n = 9$ case results in Kelvin--Voigt parameters closely matching the $n = 36$ case, with the cost function decrement $\testF$ decreasing from 2.66 to 0.20 when the constitutive model is advanced from Kelvin--Voigt to SLS. 
In contrast, the $n = 3$ case led to calibrated Kelvin--Voigt parameters with relative errors of $7\%$ and $37\%$, respectively for $G$ and $\mu$. 
The minimized cost function $\psi_0$ decreases sharply from -5.06 to -31.18 when the relaxation time scale $\tau_1$ is considered. 
This is due to the fact that exactly three experiments are considered to calibrate the three-parameter SLS model. 
Perhaps, a different subset of synthetic experiments with $n = 3$ would have resulted in a more accurate calibration of the viscoelastic model. 
However, such optimization of $\{\Rmax,\Lambda_{\max}\}$ is not feasible for real LIC experiments. 
As a general guideline, a large sample size of LIC experiments is beneficial for the performance of pIMR. 

In our LIC experiment, the measurement of collapse time has a relative accuracy on the order of $0.1\%$. 
To examine the effect of such measurement uncertainty on pIMR, we repeat the above analysis with a relative error of $0.1\%$ uniformly added to the collapse time of each experiment. 
Overall, the accuracy of the calibrated model parameters suffered minimally from the artificial error. 
In fact, for the $n = 36$ and $n = 9$ cases, the artificially increased collapse time led to a decreased $G$ in the pIMR solution, matching the input value better than in the earlier, error-free case. 
When the artificial error is further increased to $1\%$, we observe that the calibrated $G$ is decreased by approximately $11\%$ compared to the error-free case, while the accuracy of $\mu$ shifted by less than $2\%$. 
This suggests that pIMR performs stably when processing collapse time data from our LIC experiments.

\begin{table*}[tb]
    \caption{Calibrated viscoelastic parameters, minimized cost function, and cost function decrement from synthetic experiments with varying levels of artificial errors.
    The input Kelvin--Voigt model parameters are $\{G = \SI{10}{\kilo\pascal}, \mu = \SI{0.10}{\pascal\second} \}$. 
    }
    \centering
    \small
    \begin{tabular}{c c c l l l l l}
        \hline 
        Artificial Error & $n$ & Model & $G$ (\SI{}{\kilo\pascal}) & $\mu$ (\SI{}{\pascal\second}) & $\tau_1$ (\SI{}{\micro\second}) & $\psi_0$ & $\testF$ \\
        \hline 
         & 36 & NH  & 5.07 & -- & -- & -2.79 & --  \\
         & & Newtonian  & -- & $\sim{0}$ & -- & -1.94 & --  \\
         & & KV  & 10.11 & 0.095 & -- & -5.50 & 2.71 \\
         & & SLS & 10.38 & 0.116 & 0.506 & -5.78 & 0.28 \\
        \cmidrule(lr){3-8}
         $0\%$ & 9 & NH  & 4.65 & -- & -- & -2.76 & --  \\
         & & Newtonian  & -- & $\sim{0}$ & -- & -1.99 & --  \\
         & & KV  & 10.11 & 0.095 & -- & -5.42 & 2.66 \\
         & & SLS & 10.10 & 0.104 & \qty{3.34E-3}{} & -5.62 & 0.20 \\
        \cmidrule(lr){3-8}
         & 3 & NH  & 5.46 & -- & -- & -2.97 & --  \\
         & & Newtonian  & -- & $\sim{0}$ & -- & -1.91 & --  \\
         & & KV  & 9.29 & 0.063 & -- & -5.06 & 2.09  \\
         & & SLS & 10.15 & 0.107 & 1.57 & -31.18 & 26.12 \\
        \cmidrule(lr){2-8}
        & 36 & NH  & 4.97 & -- & -- & -2.79 & --  \\
         & & Newtonian  & -- & $\sim{0}$ & -- & -1.95 & --  \\
         & & KV  & 10.00 & 0.094 & -- & -5.48 & 2.69 \\
         & & SLS & 10.27 & 0.116 & 0.515 & -5.75 & 0.27 \\
        \cmidrule(lr){3-8}
         $+0.1\%$ & 9 & NH  & 4.55 & -- & -- & -2.76 & --  \\
         & & Newtonian  & -- & $\sim{0}$ & -- & -2.01 & --  \\
         & & KV  & 10.00 & 0.095 & -- & -5.39 & 2.71 \\
         & & SLS & 10.34 & 0.118 & 0.603 & -5.78 & 0.39 \\
        \cmidrule(lr){3-8}
         & 3 & NH  & 5.36 & -- & -- & -2.96 & --  \\
         & & Newtonian  & -- & $\sim{0}$ & -- & -1.92 & --  \\
         & & KV  & 9.18 & 0.063 & -- & -5.05 & 2.09 \\
         & & SLS & 10.05 & 0.107 & 1.583 & -31.78 & 26.73 \\
        \cmidrule(lr){2-8}
        & 36 & NH  & 4.07 & -- & -- & -2.79 & --  \\
         & & Newtonian  & -- & $\sim{0}$ & -- & -2.09 & --  \\
         & & KV  & 9.02 & 0.094 & -- & -5.27 & 2.48 \\
         & & SLS & 9.34 & 0.116 & 0.596 & -5.48 & 0.21 \\
        \cmidrule(lr){3-8}
         $+1\%$ & 9 & NH  & 3.65 & -- & -- & -2.76 & --  \\
         & & Newtonian  & -- & $\sim{0}$ & -- & -2.15 & --  \\
         & & KV  & 9.02 & 0.094 & -- & -5.21 & 2.45 \\
         & & SLS & 9.01 & 0.102 & \qty{7.47E-4}{} & -5.35 & 0.14 \\
        \cmidrule(lr){3-8}
         & 3 & NH  & 4.45 & -- & -- & -2.97 & --  \\
         & & Newtonian  & -- & $\sim{0}$ & -- & -2.05 & --  \\
         & & KV  & 8.18 & 0.062 & -- & -4.97 &  2.00 \\
         & & SLS & 9.12 & 0.109 & 1.686 & -31.78 & 26.81 \\
        \hline 
   \end{tabular}
   \label{tab:extra-synthetic-fit}
\end{table*}

\end{document}